\documentclass[a4paper]{article}
\usepackage{geometry}
\usepackage{amsmath}
\usepackage{amsfonts}
\usepackage{amsmath,amssymb}
\usepackage{graphicx}
\usepackage{color}

\newtheorem{theorem}{Theorem}
\newtheorem{lemma}{Lemma}
\newtheorem{corollary}{Corollary}

\newtheorem{definition}{Definition}

\newtheorem{remark}{Remark}

\newtheorem{example}{Example}

\newcommand{\ls}[1]
    {\dimen0=\fontdimen6\the\font\lineskip=#1\dimen0
     \advance\lineskip.5\fontdimen5\the\font
     \advance\lineskip-\dimen0
     \lineskiplimit=0.9\lineskip
     \baselineskip=\lineskip
     \advance\baselineskip\dimen0
     \normallineskip\lineskip\normallineskiplimit\lineskiplimit
     \normalbaselineskip\baselineskip
     \ignorespaces}

\begin{document}

\bibliographystyle{abbrv}

\title{Research on the Construction of Maximum Distance Separable Codes via Arbitrary twisted Generalized Reed-Solomon Codes}
\author{Chun'e Zhao$^{1}$, Wenping Ma$^{2}$, Tongjiang Yan$^{1}$, Yuhua Sun$^{1}$\\
$^1$ College of Sciences,
China University of Petroleum,\\
Qingdao 266555,
Shandong, China\\
$^2$ School of Telecommunication Engineering,\\
 Xidian University,
  Xi'an, China\\
Email: zhaochune1981@163.com;\ wp\_ma@mail.xidian.edu.cn;\\
 yantoji@163.com;\ sunyuhua\_1@163.com\\
}
%\date{}
 \maketitle
\footnotetext[1] {This work is supported by Shandong Provincial Natural Science Foundation of China (No. ZR2023LLZ013, No. ZR2022MA061), the Fundamental Research Funds for the Central Universities (No.22CX03015A, No.23CX03003A).
}
\thispagestyle{plain} \setcounter{page}{1}
\begin{abstract}
Maximum distance separable (MDS) codes have significant combinatorial and cryptographic applications due to their certain optimality. Generalized Reed-Solomon (GRS) codes are the most prominent MDS codes. Twisted generalized Reed-Solomon (TGRS) codes may not necessarily be MDS. It is meaningful to study the conditions under which TGRS codes are MDS. In this paper, we study a general class of TGRS (A-TGRS) codes which include all the known special ones. First, we obtain a new explicit expression of the inverse of the Vandermonde matrix. Based on this, we further derive an equivalent condition under which an A-TGRS code is MDS. According to this, the A-TGRS MDS codes include nearly all the known related results in the previous literatures. More importantly, we also obtain many other classes of MDS TGRS codes with new parameter matrices. In addition, we present a new method to compute the inverse of the lower triangular Toplitz matrix by a linear feedback shift register, which will be very useful in many research fields.

\end{abstract}

\ls{1.5}
\section{INTRODUCTION}\label{section 1}
Let $\mathbb{F}_{q}$ be a finite filed of size $q$, where $q$ is a prime power and $\mathbb{F}_{q}^{*}=\mathbb{F}_{q}\backslash \{0\}$. An $[n, k, d]$ linear code $C$ is a subspace of the linear space $\mathbb{F}_ {q}^{n}$ over $\mathbb{F}_ {q}$ with dimension $k$ and minimum Hamming distance $d$. It is well known that the parameters of $C$ satisfy $d\leq n-k+1$. If $d=n-k+1$, $C$ is called maximum distance separable (MDS). Generalized Reed-Solomon (GRS) codes are an important class of MDS linear codes. By adding some monomials (called twists) to different positions (called hooks) of each codeword polynomial, a GRS code can be generalized to twisted GRS (TGRS) code. However, a TGRS code is not necessarily MDS. After Beelen et al. firstly introduced the notion of TGRS code in 2017 \cite {8}, many researchers began to investigate the condition under which a TGRS code is MDS. Generally, there are two types of TGRS codes: single-twist and multi-twists. A TGRS code with single-twist is obtained by adding one monomial to each code polynomial, and the one with multi-twists is obtained by adding $l$ monomials to each code polynomial ($l\geq2$).

For the TGRS codes with single-twist, Beleen et al. presented a sufficient and necessary condition that a TGRS code is MDS \cite{8}. Zhang et al. studied the minimum distance and the dual code of the TGRS code for $(t,h)=(q-k-1,k-1)$ \cite{4}, where $t$ is the position of the twist and $h$ is the position of the hook. Recently, by computing all the minors of the generator matrix of $C$, Huang et al. studied the TGRS code for $(t,h)=(1,k-1)$ \cite{6}. Huang et al. and Sui et al. characterized a sufficient and necessary condition that a TGRS code is MDS for any position pair $(t,h)$ in\cite{2,3}, respectively.

  For the condition under which a TGRS code with multi-twists is MDS, Beleen et al. gave a sufficient condition for the codes constructed by adding $l$ monomials to each code polynomial \cite{9}. Zhu et al. provided a sufficient and necessary condition for double-twist TGRS codes constructed by adding one polynomial which contains no more than two monomials to each code polynomial \cite{15}. Sui et al. characterized a sufficient and necessary condition that a TGRS code constructed by adding two polynomials both of which contain no more than two monomials to each code polynomial \cite{5,1}. Gu et al. studied a sufficient and necessary condition for the ones constructed by adding $l$ special monomials to each code polynomial\cite{7}. Harshdeep et al. found a necessary and sufficient condition for the ones constructed by adding $l$ arbitrary monomials to each code polynomial\cite{10-m}.

 In short, the research listed above only studied the TGRS code for the case of special position pairs of $(t,h)$. Based on such a fact, we study the TGRS code for arbitrary position pairs of $(t,h)$, which is called an arbitrary twists Reed-Solomon (A-TGRS) code. We obtain a sufficient and necessary condition under which an A-TGRS code is MDS by determining the explicit inverse of the Vandermonde matrix. Our result shows that all the MDS TGRS codes listed above are special cases of the A-TGRS codes in this paper and their conditions such that TGRS codes are MDS are also special cases of ours. For the method to calculate the inverse of the Vandermonde matrix, early in 1967, Althaus et al. provided a method using the transition matrix between two sets of bases of a linear space, but he only mentioned this idea and did not give a specific expression \cite{18}. Until 2023, Arafat et al. gave a specific expression \cite{17}. In this paper, we provide another form of the expression using the method mentioned in \cite{18}. Based on this, we provide a new method to determine the inverse of a lower triangular Toplitz matrix which has many important applications in related research fields.

This paper is organized as follows. In Section $\mathrm{\uppercase\expandafter{\romannumeral2}}$, we recall some basic notations and symbols about A-TGRS codes. In Section $\mathrm{\uppercase\expandafter{\romannumeral3}}$, we give new explicit expressions of the inverse of the Vandermonde matrix and the inverse of a lower triangular Toplitz matrix. In Section $\mathrm{\uppercase\expandafter{\romannumeral4}}$, we give a sufficient and necessary condition under which the A-TGRS codes are MDS and obtain some new MDS TGRS codes. Moreover, employing the inverses of Toplitz matrixes, the check matrix of the TGRS code can be simplified. Finally, we conclude our work in Section $\mathrm{\uppercase\expandafter{\romannumeral5}}$.

\section{Preliminaries}\label{section 2}
Let $\mathbb{F}_{q}[x]$ be the polynomial ring over $\mathbb{F}_{q}$. Denote $\mathbb{F}_{q}[x]_{n}=\{a_{n-1}x^{n-1}+\cdots+a_{1}x+a_{0}|a_{i}\in \mathbb{F}_{q},\  i=0,1,\cdots,n-1\}$. Suppose that $\mathbf{\alpha}=(\alpha_{1},\alpha_{2},\cdots,\alpha_{n})\in\mathbb{F}_{q}^{n}$ and $v=(v_{1},v_{2},\cdots,v_{n})\in\left(\mathbb{F}_{q}^{*}\right)^{n}$, where $\alpha_{1},\alpha_{2},\cdots,\alpha_{n}$ are distinct elements. The evaluation map associated with $\mathbf{\alpha}$ and $v$ is defined as
\begin{eqnarray*}
ev_{\mathbf{\alpha,v}}&:&\mathbb{F}_{q}[x]\rightarrow \mathbb{F}_{q}^{n},\\
&&f(x)\mapsto ev_{\mathbf{\alpha,v}}(f):=(v_{1}f(\alpha_{1}),v_{2}f(\alpha_{2}),\cdots,v_{n}f(\alpha_{n})).
\end{eqnarray*}

An $[n,k]$ generalized Reed-Solomon code $GRS_{k}(\alpha,v)$ associated with $\alpha$ and $v$ is defined as
  $$GRS_{k}(\alpha,v):=\{ev_{\mathbf{\alpha,v}}(f(x)):f(x)\in \mathbb{F}_{q}[x]_{k}\}.$$ After adding  some polynomials into different positions of each $f(x)$ in $\mathbb{F}_{q}[x]_{k}$, the GRS code can be generalized as follows:

  \begin{definition}
For a positive integer $ k\leq n\leq q$ and a $k\times(n-k)$ matrix
$$A(\eta)=\left(\begin{array}{cccc}
 \eta_{0,1}&\eta_{0,2}& \cdots& \eta_{0,n-k}\\
\eta_{1,1} &  \eta_{1,2}&\cdots & \eta_{1,n-k} \\
\vdots& \vdots& & \vdots\\
 \eta_{k-1,1}& \eta_{k-1,2}&\cdots & \eta_{k-1,n-1}
 \end{array}
 \right)$$
over $\mathbb{F}_{q}$, the polynomial set
$$S=\left\{\sum\limits_{i=0}^{k-1}f_{i}x^{i}+\sum\limits_{i=0}^{k-1}f_{i}\sum\limits_{j=1}^{n-k}\eta_{i,j}x^{k-1+j}: \mathrm{for \ all\ } f_{i}\in \mathbb{F}_{q},0\leq i\leq k-1 \right\}$$
is a $k$-dimensional subspace of $\mathbb{F}_{q}[x]$. The linear code
$$C=\left\{ev_{\alpha,v}(f(x)):f(x)\in S\right\}$$
is called an arbitrary twists generalized Reed-Solomon (A-TGRS) code.
\end{definition}

In fact, the code $C$ has the following generator matrix:
\begin{equation}\label{generator matrix 1}
G=\left(\begin{array}{ccc}
1+\sum\limits_{i=1}^{n-k}\eta_{0,i}\alpha_{1}^{k-1+i} & \cdots&1+\sum\limits_{i=1}^{n-k}\eta_{0,i}\alpha_{n}^{k-1+i}\\
\alpha_{1}+\sum\limits_{i=1}^{n-k}\eta_{1,i}\alpha_{1}^{k-1+i}&\cdots &\alpha_{n}+\sum\limits_{i=1}^{n-k}\eta_{1,i}\alpha_{n}^{k-1+i}\\
 \vdots&  &\vdots\\
   \alpha_{1}^{k-1}+\sum\limits_{i=1}^{n-k}\eta_{k-1,i}\alpha_{1}^{k-1+i} &\cdots &\alpha_{n}^{k-1}+\sum\limits_{i=1}^{n-k}\eta_{k-1,i}\alpha_{1}^{k-1+i}
\end{array}
\right)\left(\begin{array}{cccc}
v_{1}&&&\\
&v_{2}$$\\
&&\ddots&\\
&&&v_{n}
\end{array}
\right).
\end{equation}

\begin{definition}
Let $k$ be a positive integer, and let $a_{0},\cdots,a_{n-1}$ be a given elements of $\mathbb{F}_{q}$. A sequence $s_{0},s_{1},\cdots$ of elements of $\mathbb{F}_{q}$ satisfying the relation
\begin{equation}\label{linear recurrence relation}
s_{t+n}=a_{n-1}s_{t+n-1}+a_{n-2}s_{t+n-2}+\cdots+a_{0}s_{t}\ for\ t=0,1,\cdots
\end{equation}
is called a ( $nth$-order ) linear feedback shift register (LFSR) sequence in $\mathbb{F}_{q}$, denoted by $(s_{k})$. The terms $s_{0},s_{1},\cdots,s_{n-1}$, which determine the rest of the sequence uniquely, are referred to as the initial values. A relation of the form (\ref{linear recurrence relation}) is called a ( $nth$-order ) linear recurrence relation.
\end{definition}

\section{The inverse of Vandermonde matrix}\label{section 3}
In order to obtain a necessary and sufficient condition such that an A-TGRS code $C$ is MDS, we need an explicit expression of the inverse of the Vandermonde matrix which is stated in the following theorem. For the inverse of the Vandermonde matrix, Arafat et al. have given an expression in \cite{17}. Here we give another form of expression which plays an important role in this paper.

\begin{theorem}\label{vandermonde matrix inverse}
Let $\alpha_{1},\alpha_{2},\cdots,\alpha_{n}$ be different elements of $\mathbb{F}_{q}^{*}$ and $G(x)=\prod\limits_{i=1}^{n}(x-\alpha_{i})=x^{n}+d_{n-1}x^{n-1}+\cdots+d_{1}x+d_{0}\in \mathbb{F}_{q}[x]$. Denote $u_{i}=\frac{1}{G'(\alpha_{i})}$ and $G_{h}(x)=\sum\limits_{j=0}^{h}d_{j}x^{j}$, where $i=1,2,\cdots,n$, and $h=0,1,\cdots,n-1$. Then the inverse of the Vandermonde matrix
$$\left(\begin{array}{cccc}
1 & \alpha_{1} &\cdots &\alpha_{1}^{n-1}\\
1 & \alpha_{2} &\cdots &\alpha_{2}^{n-1}\\
\vdots & \vdots &\ddots &\vdots\\
1 & \alpha_{n} &\cdots &\alpha_{n}^{n-1}
\end{array}
\right)^{-1}=-\left(\begin{array}{cccc}
 \frac{u_{1}}{\alpha_{1}}G_{0}(\alpha_{1})&\frac{u_{2}}{\alpha_{2}}G_{0}(\alpha_{2})  &\cdots &\frac{u_{n}}{\alpha_{n}}G_{0}(\alpha_{n})\\
 \frac{u_{1}}{\alpha_{1}^{2}}G_{1}(\alpha_{1})&\frac{u_{2}}{\alpha_{2}^{2}}G_{1}(\alpha_{2})  &\cdots &\frac{u_{n}}{\alpha_{n}^{2}}G_{1}(\alpha_{n})\\
\vdots & \vdots &\ddots &\vdots\\
 \frac{u_{1}}{\alpha_{1}^{n}}G_{n-1}(\alpha_{1})&\frac{u_{2}}{\alpha_{2}^{n}}G_{n-1}(\alpha_{2})  &\cdots &\frac{u_{n}}{\alpha_{n}^{n}}G_{n-1}(\alpha_{n})
\end{array}
\right).$$
\end{theorem}
$\mathbf{Proof.}$ Let $$f_{i}(x)=\frac{(x-\alpha_{1})(x-\alpha_{2})\cdots(x-\alpha_{i-1})(x-\alpha_{i+1})\cdots(x-\alpha_{n})}{(\alpha_{i}-\alpha_{1})(\alpha_{i}-\alpha_{2})\cdots(\alpha_{i}-\alpha_{i-1})(\alpha_{i}-\alpha_{i+1})\cdots(\alpha_{i}-\alpha_{n})},$$
where $i=1,2,\cdots,n$. Then it is not difficult to verify that $f_{1}(x),f_{2}(x),\cdots,f_{n}(x)$ is a basis of $\mathbb{F}_{q}[x]_{n}$
and that  $$f(x)=f(\alpha_{1})f_{1}(x)+f(\alpha_{2})f_{2}(x)+\cdots+f(\alpha_{n})f_{n}(x)$$ for any $f(x)\in \mathbb{F}_{q}[x]_{n}$.
By the basic theory of linear algebra we know that the transitional matrix from the basis $$f_{1}(x),f_{2}(x),\cdots,f_{n}(x) \mathrm{\ \ \ to }\  \ \ 1,x,\cdots,x^{n-1}$$ is exactly the Vandermonde matrix
$$V=\left(\begin{array}{cccc}
1 & \alpha_{1} &\cdots &\alpha_{1}^{n-1}\\
1 & \alpha_{2} &\cdots &\alpha_{2}^{n-1}\\
\vdots & \vdots &\ddots &\vdots\\
1 & \alpha_{n} &\cdots &\alpha_{n}^{n-1}
\end{array}
\right),$$
i.e., $(f_{1}(x),f_{2}(x),\cdots,f_{n}(x))=(1,x,\cdots,x^{n-1})V^{-1}$.
Without loss of generality, denote $$V^{-1}=\left(\begin{array}{cccc}
v_{1,0} & v_{2,0} &\cdots &v_{n,0}\\
v_{1,1} &v_{2,1} &\cdots &v_{n,1}\\
\vdots & \vdots &\ddots &\vdots\\
v_{1,n-1} &v_{2,n-1} &\cdots &v_{n,n-1}
\end{array}
\right).$$
Then the $i$th column of $V^{-1}$ is the coordinate of $f_{i}(x)$ under the base $1,x,\cdots,x^{n-1}$, i.e., $f_{i}(x)=\sum\limits_{k=0}^{n-1}v_{i,k}x^{k}, i=1,2,\cdots,n$.
Thus we have
 \begin{equation}\label{polynomial}
 f_{i}(x)(x-\alpha_{i})=u_{i}G(x).
 \end{equation}
 Comparing the corresponding coefficients of $x^{h},x^{h-1},\cdots,x,1$ in the two sides of Eq. (\ref{polynomial}), we have the following equations
$$\left\{\begin{array}{cc}
v_{i,h-1}-\alpha_{i}v_{i,h}&=u_{i}d_{h},\\
v_{i,h-2}-\alpha_{i}v_{i,h-1}&=u_{i}d_{h-1},\\
\cdots  ,\\
v_{i,0}-\alpha_{i}v_{i,1}&=u_{i}d_{1},\\
-\alpha_{i}v_{i,0}&=u_{i}d_{0} .
\end{array}\right.$$
For convenience, denote the equation with $u_{i}d_{k}$ being on its right in the above by $(k)$. Then by calculating the sum $\sum\limits_{k=0}^{h}(k)\ast \alpha_{i}^{k}$, we have the following equation
$$-v_{i,h}\alpha_{i}^{h+1}=u_{i}[d_{h}\alpha_{i}^{h}+d_{h-1}\alpha_{i}^{h-1}+\cdots+d_{1}\alpha_{i}+d_{0}]=u_{i}G_{h}(\alpha_{i}),$$
which implies $$v_{i,h}=-\frac{u_{i}}{\alpha_{i}^{h+1}}G_{h}(\alpha_{i}),$$ where $i=1,2,\cdots,n$ and $h=0,1,2,\cdots,n-1$.

Using Theorem \ref{vandermonde matrix inverse}, it is easy to deduce Theorem 3 in \cite{17}. In addition, we can obtain the following results correspondingly.

\begin{corollary}\label{expression of Vandermonde matrix}
Let $\alpha_{1},\alpha_{2},\cdots,\alpha_{n}$ be different elements in $\mathbb{F}_{q}^{*}$ and $G(x)=\prod\limits_{i=1}^{n}(x-\alpha_{i})=x^{n}+d_{n-1}x^{n-1}+\cdots+d_{1}x+d_{0}\in \mathbb{F}_{q}[x]$. Set $u_{i}=\frac{1}{G'(\alpha_{i})}$, $i=1,2,\cdots,n$. Then the inverse of the Vandermonde matrix is
$$\left(\begin{array}{cccc}
1 & \alpha_{1} &\cdots &\alpha_{1}^{n-1}\\
1 & \alpha_{2} &\cdots &\alpha_{2}^{n-1}\\
\vdots & \vdots &\ddots &\vdots\\
1 & \alpha_{n} &\cdots &\alpha_{n}^{n-1}
\end{array}
\right)^{-1}=-\left(\begin{array}{cccc}
d_{0} & 0 & \cdots&0\\
 d_{1}&d_{0}&\cdots &0\\
 \vdots&  \vdots&\ddots &\vdots\\
 d_{n-1}& d_{n-2} &\cdots &d_{0}
\end{array}
\right)\left(\begin{array}{cccc}
\frac{1}{\alpha_{1}} &\frac{1}{\alpha_{2}}  &\cdots &\frac{1}{\alpha_{n}}\\
\frac{1}{\alpha_{1}^{2}} & \frac{1}{\alpha_{2}^{2}} &\cdots &\frac{1}{\alpha_{n}^{2}}\\
 \vdots& \vdots &\ddots &\vdots\\
\frac{1}{\alpha_{1}^{n}} &\frac{1}{\alpha_{2}^{n}}  & \cdots&\frac{1}{\alpha_{n}^{n}}
\end{array}
\right)\left(\begin{array}{cccc}
u_{1} &  & &\\
 & u_{2} & &\\
 &  &\ddots &\\
 &  & &u_{n}
\end{array}
\right) .$$
\end{corollary}

\begin{remark}
Let $\alpha_{1},\alpha_{2},\cdots,\alpha_{n}$ be different elements in $\mathbb{F}_{q}$, $G(x)=\prod\limits_{i=1}^{n}(x-\alpha_{i})$, $u_{i}=\frac{1}{G'(\alpha_{i})}$, $i=1,2,\cdots,n$ and $w_{t}=\sum\limits_{i=1}^{n}u_{i}\alpha_{i}^{t}$ for any integer $t$. Then there is an important relation between $w_{t}$ and the determinant of the matrix with the form
 $$\left(\begin{array}{cccc}
1 & 1 &\cdots &1\\
\alpha_{1} & \alpha_{2} &\cdots &\alpha_{n}\\
\vdots & \vdots & &\vdots\\
\alpha_{1}^{n-2} & \alpha_{2}^{n-2}&\cdots &\alpha_{n}^{n-2}\\
\alpha_{1}^{t} & \alpha_{2}^{t} &\cdots &\alpha_{n}^{t}
\end{array} \right).$$
We list it in the following theorem.
\end{remark}

\begin{theorem}\label{wk and diagram}
Let $\alpha_{1},\alpha_{2},\cdots,\alpha_{n}$ be different elements in finite field $\mathbb{F}_{q}$ and $G(x)=\prod\limits_{i=1}^{n}(x-\alpha_{i})=x^{n}+d_{n-1}x^{n-1}+\cdots+d_{1}x+d_{0}\in \mathbb{F}_{q}[x]$. Suppose $u_{i}=\frac{1}{G'(\alpha_{i})},\ i=1,2,\cdots,n$. Then for any integer $t$, we have
$$\frac{1}{\Delta}\left|\begin{array}{cccc}
1 & 1 &\cdots &1\\
\alpha_{1} & \alpha_{2} &\cdots &\alpha_{n}\\
\vdots & \vdots & &\vdots\\
\alpha_{1}^{n-2} & \alpha_{2}^{n-2}&\cdots &\alpha_{n}^{n-2}\\
\alpha_{1}^{t} & \alpha_{2}^{t} &\cdots &\alpha_{n}^{t}
\end{array} \right|=w_{t},$$
 where $w_{t}=\sum\limits_{i=1}^{n}u_{i}\alpha_{i}^{t}$ and $\Delta=\prod\limits_{1\leq i<j\leq n}(\alpha_{j}-\alpha_{i})$.
\end{theorem}
$\mathbf{Proof.}$
Let $D_{i}$ be the algebraic cofactor of the element $\alpha_{i}^{n-1}$  in the last row of the following determinant,
$$\left|\begin{array}{ccccc}
1 & 1 &1&\cdots &1\\
\alpha_{1} & \alpha_{2}&\alpha_{3} &\cdots &\alpha_{n}\\
\vdots & \vdots &\vdots&\ddots &\vdots\\
\alpha_{1}^{n-1} & \alpha_{2}^{n-1} &\alpha_{3}^{n-1}&\cdots &\alpha_{n}^{n-1}
\end{array} \right|,$$
 $i=1,2,\cdots,n$.

Then
$$\left\{\begin{array}{cc}
D_{1}+D_{2}+\cdots+D_{n} &=0 ,\\
\alpha_{1}D_{1}+\alpha_{2}D_{2}+\cdots+\alpha_{n}D_{n} &=0,\\
\vdots&\\
 \alpha_{1}^{n-2}D_{1}+\alpha_{2}^{n-2}D_{2}+\cdots+\alpha_{n}^{n-2}D_{n} &=0.
\end{array}\right.$$
So $(D_{1},D_{2},\cdots,D_{n})^{T}$ is a solution of the following equation system
$$\left\{\begin{array}{cc}
x_{1}+x_{2}+\cdots+x_{n} &=0 ,\\
\alpha_{1}x_{1}+\alpha_{2}x_{2}+\cdots+\alpha_{n}x_{n} &=0,\\
\vdots&\\
 \alpha_{1}^{n-2}x_{1}+\alpha_{2}^{n-2}x_{2}+\cdots+\alpha_{n}^{n-2}x_{n} &=0.
\end{array}\right.$$
Therefore $(\frac{D_{1}}{\Delta},\frac{D_{2}}{\Delta},\cdots,\frac{D_{1}}{\Delta})^{T}$ is also a solution of the above equation system.

By direct calculation, we get $\frac{D_{i}}{\Delta}=u_{i},i=1,2,\cdots,n$.
Thus
$$\frac{1}{\Delta}\left|\begin{array}{ccccc}
1 & 1 &1&\cdots &1\\
\alpha_{1} & \alpha_{2}&\alpha_{3} &\cdots &\alpha_{n}\\
\vdots & \vdots &\vdots&\ddots &\vdots\\
\alpha_{1}^{n-2} & \alpha_{2}^{n-2} &\alpha_{3}^{n-2}&\cdots &\alpha_{n}^{n-2}\\
\alpha_{1}^{t} & \alpha_{2}^{t} &\alpha_{3}^{t}&\cdots &\alpha_{n}^{t}
\end{array} \right|=\frac{1}{\Delta}\sum\limits_{i=1}^{n}\alpha_{i}^{t}D_{i}=\sum\limits_{i=1}^{n}\alpha_{i}^{t}u_{i}=w_{t}.$$

According to Theorem \ref{wk and diagram}, it is easy to get the following result which is important in the later discussion.

\begin{corollary}\label{zero and wk}
Let $\alpha_{1},\alpha_{2},\cdots,\alpha_{n}$ be different elements in $\mathbb{F}_{q}$ and $\alpha_{1}=0$, $G(x)=\prod\limits_{i=1}^{n}(x-\alpha_{i})$, $u_{j}=\frac{1}{G'(\alpha_{j})}$, $w_{t}=\sum\limits_{j=1}^{n}\alpha_{j}^{t}u_{j}$, $g(x)=\prod\limits_{i=2}^{n}(x-\alpha_{i})$, $u'_{i}=\frac{1}{g'(\alpha_{i})}$,$w'_{t}=\sum\limits_{i=2}^{n}\alpha_{i}^{t}u'_{i}$.
Then
$$w'_{t}=w_{t+1}$$
for any integer $t$.
\end{corollary}
$\mathbf{Proof.}$
Let $\Delta=\prod\limits_{1\leq i<j\leq n}(\alpha_{j}-\alpha_{i})$ and $\Delta_{1}=\prod\limits_{2\leq i<j\leq n}(\alpha_{j}-\alpha_{i})$.
By Theorem \ref{wk and diagram}, then for any integer $t$, we have $$w_{t+1}=\frac{1}{\Delta}\left|\begin{array}{cccc}
1&1&\cdots&1\\
0&\alpha_{2}&\cdots&\alpha_{n}\\
\vdots&\vdots&&\vdots\\
0&\alpha_{2}^{n-2}&\cdots&\alpha_{n}^{n-2}\\
0&\alpha_{2}^{t+1}&\cdots&\alpha_{n}^{t+1}
\end{array}\right|=\frac{1}{\Delta}\left|\begin{array}{ccc}
\alpha_{2}&\cdots&\alpha_{n}\\
\vdots&&\vdots\\
\alpha_{2}^{n-2}&\cdots&\alpha_{n}^{n-2}\\
\alpha_{2}^{t+1}&\cdots&\alpha_{n}^{t+1}
\end{array}\right|=\frac{\prod\limits_{i=2}^{n}\alpha_{i}}{\Delta}\frac{1}{\Delta_{1}}w'_{t}=w'_{t}.$$
\begin{theorem}\label{recursive formulation of wk}
Let $\alpha_{1},\alpha_{2},\cdots,\alpha_{n}$ be different elements in $\mathbb{F}_{q}$ and $G(x)=\prod\limits_{i=1}^{n}(x-\alpha_{i})=x^{n}+d_{n-1}x^{n-1}+\cdots+d_{1}x+d_{0}\in \mathbb{F}_{q}[x]$. Suppose $u_{i}=\frac{1}{G'(\alpha_{i})}$ and $w_{t}=\sum\limits_{i=1}^{n}u_{i}\alpha_{i}^{t}$, where $i=1,2,\cdots,n$. Then $w_{0}=w_{1}=\cdots=w_{n-2}=0$, $w_{n-1}=1$, and for any integer $t$, the linear recurrence relation is :
\begin{equation}\label{recursive formulation}
w_{t}=-(d_{n-1}w_{t-1}+d_{n-2}w_{t-2}+\cdots+d_{0}w_{t-n}).
\end{equation}
\end{theorem}

$\mathbf{Proof.}$
From Theorem \ref{wk and diagram}, it is easy to see that $w_{0}=w_{1}=\cdots=w_{n-2}=0,w_{n-1}=1$. Next we only need to prove Eq. (\ref{recursive formulation}).
Suppose
\begin{equation}\label{kleim fuction}
\left(\begin{array}{cccc}
1 & \alpha_{1} &\cdots &\alpha_{1}^{n-1}\\
1 & \alpha_{2} &\cdots &\alpha_{2}^{n-1}\\
\vdots & \vdots &\ddots &\vdots\\
1 & \alpha_{n} &\cdots &\alpha_{n}^{n-1}
\end{array}
\right)\left(\begin{array}{c}
x_{0} \\
x_{1} \\
\vdots \\
x_{n-1}
\end{array}
\right)=\left(\begin{array}{c}
\alpha_{1}^{t} \\
\alpha_{2}^{t} \\
\vdots \\
\alpha_{n}^{t}
\end{array}
\right).
\end{equation}
Then $\left(\begin{array}{c}
x_{0} \\
x_{1} \\
\vdots \\
x_{n-1}
\end{array}
\right)=\left(\begin{array}{cccc}
1 & \alpha_{1} &\cdots &\alpha_{1}^{n-1}\\
1 & \alpha_{2} &\cdots &\alpha_{2}^{n-1}\\
\vdots & \vdots &\ddots &\vdots\\
1 & \alpha_{n} &\cdots &\alpha_{n}^{n-1}
\end{array}
\right)^{-1}\left(\begin{array}{c}
\alpha_{1}^{t} \\
\alpha_{2}^{t} \\
\vdots \\
\alpha_{n}^{t}
\end{array}
\right).$\\
\\
We will discuss the recurrence relation for two cases.

(1) $\alpha_{1},\alpha_{2},\cdots,\alpha_{n}\in \mathbb{F}_{q}^{*}$.

On the one hand, combining Theorem \ref{vandermonde matrix inverse} and Corollary \ref{expression of Vandermonde matrix}, we have
\begin{eqnarray*}
x_{n-1}&=&-(d_{n-1},d_{n-2},\cdots,d_{0})\left(\begin{array}{cccc}
\frac{1}{\alpha_{1}} & \frac{1}{\alpha_{2}} &\cdots &\frac{1}{\alpha_{n}}\\
\frac{1}{\alpha_{1}^{2}} & \frac{1}{\alpha_{2}^{2}} &\cdots &\frac{1}{\alpha_{n}^{2}}\\
\vdots & \vdots &\ddots &\vdots\\
\frac{1}{\alpha_{1}^{n}} & \frac{1}{\alpha_{2}^{n}} &\cdots &\frac{1}{\alpha_{n}^{n}}
\end{array}
\right)\left(\begin{array}{cccc}
u_{1} &  & &\\
 & u_{2} & &\\
 &  &\ddots &\\
 &  & &u_{n}
\end{array}
\right)\left(\begin{array}{c}
\alpha_{1}^{t} \\
\alpha_{2}^{t} \\
\vdots \\
\alpha_{n}^{t}
\end{array}
\right)\\
&=&-(d_{n-1},d_{n-2},\cdots,d_{0})\left(\begin{array}{c}
 u_{1}\alpha_{1}^{t-1}+u_{2}\alpha_{2}^{t-1}+\cdots+u_{n}\alpha_{n}^{t-1}\\
 u_{1}\alpha_{1}^{t-2}+u_{2}\alpha_{2}^{t-2}+\cdots+u_{n}\alpha_{n}^{t-2}\\
\vdots \\
u_{1}\alpha_{1}^{t-n}+u_{2}\alpha_{2}^{t-n}+\cdots+u_{n}\alpha_{n}^{t-n}\\
\end{array}
\right)\\
&=&-(d_{n-1},d_{n-2},\cdots,d_{0})\left(\begin{array}{c}
w_{t-1}\\
w_{t-2}\\
\vdots \\
w_{t-n}\\
\end{array}
\right)\\
&=&-(d_{n-1}w_{t-1}+d_{n-2}w_{t-2}+\cdots+d_{0}w_{t-n}).
\end{eqnarray*}

On the other hand, according to Cramer's Rule, we also have

$$x_{n-1}=\frac{\left|\begin{array}{cccc}
1 & 1 &\cdots &1\\
\alpha_{1} & \alpha_{2} &\cdots &\alpha_{n}\\
\vdots & \vdots & &\vdots\\
\alpha_{1}^{n-2} & \alpha_{2}^{n-2}&\cdots &\alpha_{n}^{n-2}\\
\alpha_{1}^{t} & \alpha_{2}^{t} &\cdots &\alpha_{n}^{t}
\end{array} \right|}{\left|\begin{array}{cccc}
1 & 1 &\cdots &1\\
\alpha_{1} & \alpha_{2} &\cdots &\alpha_{n}\\
\vdots & \vdots & &\vdots\\
\alpha_{1}^{n-2} & \alpha_{2}^{n-2}&\cdots &\alpha_{n}^{n-2}\\
\alpha_{1}^{n-1} & \alpha_{2}^{n-1} &\cdots &\alpha_{n}^{n-1}
\end{array} \right|}.$$
By Theorem \ref{wk and diagram}, $x_{n-1}=w_{t}$. So
$$w_{t}=-(d_{n-1}w_{t-1}+d_{n-2}w_{t-2}+\cdots+d_{0}w_{t-n}).$$
(2) $0\in \{\alpha_{1},\cdots,\alpha_{n}\}$.

Without loss of generality, let $\alpha_{1}=0$. Then $G(x)=\prod\limits_{i=1}^{n}(x-\alpha_{i})=x^{n}+d_{n-1}x^{n-1}+\cdots+d_{1}x+d_{0}$ with $d_{0}=0$. So $G(x)=xg(x)$, where $g(x)=\prod\limits_{i=2}^{n}(x-\alpha_{i})$. Let $u'_{i}=\frac{1}{g'(\alpha_{i})},i=2,\cdots,n$ and $w'_{t}=\sum\limits_{i=2}^{n}\alpha_{i}^{t}u'_{i}$. Then by the proof of (1), it can be seen that
$w'_{t-1}=-(d_{n-1}w'_{t-2}+d_{n-2}w'_{t-3}+\cdots+d_{1}w'_{t-n}).$ By Corollary \ref{zero and wk}, we have

$w_{t}=-(d_{n-1}w_{t-1}+d_{n-2}w_{t-2}+\cdots+d_{1}w_{t-n+1})=-(d_{n-1}w_{t-1}+d_{n-2}w_{t-2}+\cdots+d_{1}w_{t-n+1}+d_{0}w_{n-t})$.

As another important application of Theorem \ref{vandermonde matrix inverse}, we can also obtain the the inverse of the Toplitz matrix which is needed in the sequel and also may be useful in many research fields.
\begin{theorem}\label{inverse of  Toplitz Matrix}
Let $\alpha_{1},\alpha_{2},\cdots,\alpha_{n}$ be different elements in $\mathbb{F}_{q}$ and $G(x)=\prod\limits_{i=1}^{n}(x-\alpha_{i})=x^{n}+c_{1}x^{n-1}+\cdots+c_{n-1}x+c_{n}$, then the inverse of a lower triangular Toplitz matrix is
\begin{equation}\label{inverse of T}
\left(\begin{array}{cccc}
1 & 0 & \cdots&0\\
 c_{1}&1&\cdots &0\\
 \vdots&  \vdots&\ddots &\vdots\\
 c_{n-1}& c_{n-2} &\cdots &1
\end{array}
\right)^{-1}=\left(\begin{array}{cccc}
w_{n-1} & 0 &\cdots &0\\
w_{n}&w_{n-1} &\cdots &0\\
\vdots & \vdots &\ddots &\vdots\\
w_{2n-2} &  w_{2n-1} &\cdots & w_{n-1}
\end{array}
\right),
\end{equation}
where $w_{i}=\sum\limits_{j=1}^{n}\alpha_{j}^{i}u_{j}$, $u_{j}=\frac{1}{G'(\alpha_{j})}$, $j=1,2,\cdots,n$ and $i$ is any integer.
\end{theorem}
$\mathbf{Proof.}$\
We will introduce the result for two cases.

 (1)\ \ $\alpha_{1},\alpha_{2},\cdots,\alpha_{n}\in \mathbb{F}_{q}^{*}$.

Let $\widetilde{G}(x)=x^{n}+\frac{c_{n-1}}{c_{n}}x^{n-1}+\cdots+\frac{c_{1}}{c_{n}}x+\frac{1}{c_{n}}$. Then all the roots of $\widetilde{G}(x)$ are exactly $$\frac{1}{\alpha_{1}},\frac{1}{\alpha_{2}},\cdots,\frac{1}{\alpha_{n}}.$$ By Corollary \ref{expression of Vandermonde matrix}, the inverse of the lower triangular matrix
$$\left(\begin{array}{cccc}
\frac{1}{c_{n}} & 0 & \cdots&0\\
\frac{c_{1}}{c_{n}}&\frac{1}{c_{n}}&\cdots &0\\
 \vdots&  \vdots&\ddots &\vdots\\
 \frac{c_{n-1}}{c_{n}}& \frac{c_{n-2}}{c_{n}} &\cdots &\frac{1}{c_{n}}
\end{array}
\right)^{-1}=-\left(\begin{array}{cccc}
\alpha_{1} &\alpha_{2}  &\cdots &\alpha_{n}\\
\alpha_{1} ^{2} &\alpha_{2}^{2} &\cdots &\alpha_{n}^{2}\\
 \vdots& \vdots &\ddots &\vdots\\
\alpha_{1}^{n} &\alpha_{2}^{n}  & \cdots&\alpha_{n}^{n}
\end{array}
\right)\left(\begin{array}{cccc}
u_{1}^{*} &  & &\\
 & u_{2}^{*} & &\\
 &  &\ddots &\\
 &  & &u_{n}^{*}
\end{array}
\right)\left(\begin{array}{cccc}
1 &\frac{1}{\alpha_{1}} &\cdots &\frac{1}{\alpha_{1}^{n-1}}\\
1 & \frac{1}{\alpha_{2}} &\cdots &\frac{1}{\alpha_{2}^{n-1}}\\
 \vdots& \vdots &\ddots &\vdots\\
1 &\frac{1}{\alpha_{n}}  & \cdots&\frac{1}{\alpha_{n}^{n-1}}
\end{array}
\right),$$
 where $u_{i}^{*}=\frac{1}{\widetilde{G}^{'}(\frac{1}{\alpha_{i}})},i=1,2,\cdots,n$. By calculation, it is easy to see $\frac{u_{i}^{*}}{c_{n}}=-u_{i}\alpha_{i}^{n-2}.$ So

 \begin{eqnarray*}
\left(\begin{array}{cccc}
1 & 0 & \cdots&0\\
 c_{1}&1&\cdots &0\\
 \vdots&  \vdots&\ddots &\vdots\\
 c_{n-1}& c_{n-2} &\cdots &1
\end{array}
\right)^{-1}&=&-\left(\begin{array}{cccc}
\alpha_{1} &\alpha_{2}  &\cdots &\alpha_{n}\\
\alpha_{1} ^{2} &\alpha_{2}^{2} &\cdots &\alpha_{n}^{2}\\
 \vdots& \vdots &\ddots &\vdots\\
\alpha_{1}^{n} &\alpha_{2}^{n}  & \cdots&\alpha_{n}^{n}
\end{array}
\right)\left(\begin{array}{cccc}
\frac{u'_{1}}{c_{n}} &  & &\\
 & \frac{u'_{2}}{c_{n}} & &\\
 &  &\ddots &\\
 &  & &\frac{u'_{n}}{c_{n}}
\end{array}
\right)\left(\begin{array}{cccc}
1 &\frac{1}{\alpha_{1}} &\cdots &\frac{1}{\alpha_{1}^{n-1}}\\
1 & \frac{1}{\alpha_{2}} &\cdots &\frac{1}{\alpha_{2}^{n-1}}\\
 \vdots& \vdots &\ddots &\vdots\\
1 &\frac{1}{\alpha_{n}}  & \cdots&\frac{1}{\alpha_{n}^{n-1}}
\end{array}
\right)\\
&=&\left(\begin{array}{cccc}
\alpha_{1} &\alpha_{2}  &\cdots &\alpha_{n}\\
\alpha_{1} ^{2} &\alpha_{2}^{2} &\cdots &\alpha_{n}^{2}\\
 \vdots& \vdots &\ddots &\vdots\\
\alpha_{1}^{n} &\alpha_{2}^{n}  & \cdots&\alpha_{n}^{n}
\end{array}
\right)\left(\begin{array}{ccc}
u_{1}\alpha_{1}^{n-2}&  &\\
    &\ddots &\\
   & &u_{n}\alpha_{n}^{n-2}
\end{array}
\right)\left(\begin{array}{cccc}
1 &\frac{1}{\alpha_{1}} &\cdots &\frac{1}{\alpha_{1}^{n-1}}\\
1 & \frac{1}{\alpha_{2}} &\cdots &\frac{1}{\alpha_{2}^{n-1}}\\
 \vdots& \vdots &\ddots &\vdots\\
1 &\frac{1}{\alpha_{n}}  & \cdots&\frac{1}{\alpha_{n}^{n-1}}
\end{array}
\right)\\
&=&\left(\begin{array}{cccc}
w_{n-1} & 0 &\cdots &0\\
w_{n}&w_{n-1} &\cdots &0\\
\vdots & \vdots &\ddots &\vdots\\
w_{2n-2} &  w_{2n-1} &\cdots & w_{n-1}
\end{array}
\right).
\end{eqnarray*}

(2) \ \ $0\in\{\alpha_{1},\alpha_{2},\cdots,\alpha_{n}\}$.

Without loss of generality, let $\alpha_{1}=0$. Then $G(x)=x^{n}+c_{1}x^{n-1}+\cdots+c_{n-1}x=xg(x)$ and $g(x)=\prod\limits_{i=2}^{n}(x-\alpha_{i})$. Let $u'_{i}=\frac{1}{g'(\alpha_{i})}$, $w'_{t}=\sum\limits_{i=2}^{n}\alpha_{i}^{t}u'_{i}$, $i=2,3,\cdots,n$.
The Toplitz Matrix is denoted by
$$\left(\begin{array}{ccccc}
1 & 0 & \cdots&0&0\\
 c_{1}&1&\cdots &0&0\\
 \vdots&  \vdots&\ddots &\vdots&0\\
 c_{n-2}& c_{n-3} &\cdots &1&0\\
 c_{n-1}& c_{n-2} &\cdots &c_{1}&1
\end{array}
\right)\triangleq\left(\begin{array}{cc}
T&0\\
\delta&1
\end{array}
\right),$$
where $T=\left(\begin{array}{cccc}
1 & 0 & \cdots&0\\
 c_{1}&1&\cdots &0\\
 \vdots&  \vdots&\ddots &\vdots\\
  c_{n-2}& c_{n-3} &\cdots &1
\end{array}
\right)$, $0=\left(\begin{array}{c}
0\\
0\\
\vdots\\
0
\end{array}
\right)$, $\delta=(c_{n-1}, c_{n-2}, \cdots, c_{1}).$

According to the method of the block matrix, it is not difficult to obtain
$$\left(\begin{array}{cc}
T&0\\
\delta&1
\end{array}
\right)^{-1}=\left(\begin{array}{cc}
T^{-1} &0 \\
-\delta T^{-1}&1
\end{array}
\right).$$
By the proof of (1) and Corollary \ref{zero and wk}, we have
\begin{align*}
T^{-1}=\left(\begin{array}{cccc}
w'_{n-2} & 0 &\cdots &0\\
w'_{n-1}&w'_{n-2} &\cdots &0\\
\vdots & \vdots &\ddots &\vdots\\
w'_{2n-4} &  w'_{2n-3} &\cdots & w'_{n-2}
\end{array}
\right)=\left(\begin{array}{cccc}
w_{n-1} & 0 &\cdots &0\\
w_{n}&w_{n-1} &\cdots &0\\
\vdots & \vdots &\ddots &\vdots\\
w_{2n-3} &  w_{2n-4} &\cdots & w_{n-1}
\end{array}
\right).
\end{align*}
By Theorems \ref{wk and diagram}-\ref{recursive formulation of wk},
\begin{align*}
-\delta T^{-1}&=-(c_{n-1},c_{n-2},\cdots,c_{1})\left(\begin{array}{cccc}
w_{n-1} & 0 &\cdots &0\\
w_{n}&w_{n-1} &\cdots &0\\
\vdots & \vdots &\ddots &\vdots\\
w_{2n-3} &  w_{2n-4} &\cdots & w_{n-1}
\end{array}
\right)\\
&=-(c_{n},c_{n-1},c_{n-2},\cdots,c_{1})\left(\begin{array}{ccccc}
0 &0&0 &\cdots &0\\
w_{n-1}&0 &0&\cdots &0\\
w_{n}&w_{n-1} &0&\cdots &0\\
\vdots & \vdots &\vdots&\ddots &\vdots\\
w_{2n-3} &  w_{2n-4} &\cdots &w_{n-2}& w_{n-1}
\end{array}
\right)\\
&=-(c_{n},c_{n-1},c_{n-2},\cdots,c_{1})\left(\begin{array}{ccccc}
w_{n-2} &0&0 &\cdots &0\\
w_{n-1}&w_{n-2} &0&\cdots &0\\
w_{n}&w_{n-1} &0&\cdots &0\\
\vdots & \vdots &\vdots&\ddots &\vdots\\
w_{2n-3} &  w_{2n-4} &\cdots &w_{n-2}& w_{n-2}\\
w_{2n-3} &  w_{2n-4} &\cdots &w_{n-2}& w_{n-1}
\end{array}
\right)\\
&=(w_{2n-2},w_{2n-3},\cdots,w_{n}).
\end{align*}
By Theorem \ref{wk and diagram}, $w_{n-1}=1$. The proof completes.

According to the properties of a matrix and its inverse, we get the following result immediately.

\begin{corollary}\label{generalized Toplitz matirx inverse}
Let $\alpha_{1},\alpha_{2},\cdots,\alpha_{n}$ be different elements in $\mathbb{F}_{q}$ and $G(x)=\prod\limits_{i=1}^{n}(x-\alpha_{i})=x^{n}+c_{1}x^{n-1}+\cdots+c_{n-1}x+c_{n}$. Suppose $c_{i}=0$ for $i>n$. For any positive integer $t$, we have
$$\left(\begin{array}{cccc}
1 & 0 & \cdots&0\\
 c_{1}&1&\cdots &0\\
 \vdots&  \vdots&\ddots &\vdots\\
 c_{t}&c_{t-1} &\cdots &1
\end{array}
\right)^{-1}=\left(\begin{array}{cccc}
w_{n-1} & 0 &\cdots &0\\
w_{n}&w_{n-1} &\cdots &0\\
\vdots & \vdots &\ddots &\vdots\\
w_{n-1+t} &  w_{n-2+t} &\cdots & w_{n-1}
\end{array}
\right),
$$
 where $w_{i}=\sum\limits_{j=1}^{n}\alpha_{j}^{i}u_{j}$, $u_{j}=\frac{1}{G^{'}(\alpha_{j})}$, \ $j=1,2,\cdots,n$ and \ $i$ is any integer.
\end{corollary}
Using the same method as that used in the proof of Theorem \ref{inverse of  Toplitz Matrix}, we also have the following result.

\begin{corollary}\label{minus Toplitz matirx inverse}
Let $\alpha_{1},\alpha_{2},\cdots,\alpha_{n}$ be different elements in $\mathbb{F}_{q}^{*}$ and $G(x)=\prod\limits_{i=1}^{n}(x-\alpha_{i})=x^{n}+c_{1}x^{n-1}+\cdots+c_{n-1}x+c_{n}$. Then we have
$$\left(\begin{array}{cccc}
c_{n} & 0 & \cdots&0\\
 c_{n-1}&c_{n}&\cdots &0\\
 \vdots&  \vdots&\ddots &\vdots\\
 c_{1}&c_{2} &\cdots &c_{n}
\end{array}
\right)^{-1}=-\left(\begin{array}{cccc}
w_{-1} & 0 &\cdots &0\\
w_{-2}&w_{-1} &\cdots &0\\
\vdots & \vdots &\ddots &\vdots\\
w_{-n} &  w_{-2} &\cdots & w_{-1}
\end{array}
\right),
$$\ where $w_{i}=\sum\limits_{j=1}^{n}\alpha_{j}^{i}u_{j}$, $u_{j}=\frac{1}{G'(\alpha_{j})}$, $j=1,2,\cdots,n$ and $i$ is any integer.
\end{corollary}

Based on the above results of Theorem \ref{inverse of  Toplitz Matrix}, Corollary \ref{generalized Toplitz matirx inverse} and Theorem \ref{recursive formulation of wk}, we now give a new method of computing the inverse of a lower triangular Toplitz matrix in the following.
For convenience, we denote the matrix $$\left(\begin{array}{ccccc}
c_{0} &  & & &\\
c_{1}&c_{0}& & &\\
c_{2}&c_{1}&c_{0} & &\\
 \vdots&\vdots&\ddots &\ddots&\\
 c_{n-1}&c_{n-2}&\cdots&c_{1}&c_{0}\\
 \end{array}
\right)\triangleq T(c_{0},c_{1},\cdots,c_{n-1}).$$
Then Eq. (\ref{inverse of T}) in Theorem \ref{inverse of  Toplitz Matrix} can be simply denoted as
$T(1,c_{1},\cdots,c_{n-1})^{-1}=T(w_{n-1},w_{n},\cdots,w_{2n-2}).$ In the following theorem, we provide a concise algorithm by theory of LFSR.
\begin{theorem}\label{State transition formula}

Let $(w_{k})$ be a LFSR sequence in $\mathbb{F}_{q}$, with linear recurrence relation $$w_{k}=-(c_{1}w_{k-1}+c_{2}w_{k-2}+\cdots+c_{n}w_{k-n}) \ \ for\ k\geq n  $$
and initial values $w_{0}=w_{1}=\cdots=w_{n-2}=0,\ w_{n-1}=1.$
Assume $G(x)=x^{n}+c_{1}x^{n-1}+\cdots+c_{n-1}x+c_{n}$ has $n$ different roots in $\mathbb{F}_{q}$. Then
$$T(1,c_{1},\cdots,c_{n-1})^{-1}=T(w_{n-1},w_{n},\cdots,w_{2n-2}),$$
i.e., each $w_{i}$ ($n-1\leq i\leq 2n-2$) in $T(w_{n-1},w_{n},\cdots,w_{2n-2})$ is exactly the corresponding element of  the sequence $(w_{k})$.

% where $(w_{k})$ be a linear feedback shift register sequence generated by as initial state, and state transition formula is
%$$\left(\begin{array}{c}
%w_{k+1}\\
%w_{k+2}\\
%\vdots \\
%w_{k+n}\\
%\end{array}
%\right) \\
%=\left(\begin{array}{cccccc}
%0 & 1 & & &\\
% & 0 & 1 &  &\\
%  &   & \ddots  &\ddots &\\
%  &   &   &     0&    1   &\\
% -c_{n}&-c_{n-1}& -c_{n-2}&\cdots&-c_{1}\\
%\end{array}
%\right)\left(\begin{array}{c}
%w_{k}\\
%w_{k+1}\\
%\vdots \\
%w_{k+n-1}\\
%\end{array}
%\right).\\
%$$
\end{theorem}

\section{MDS condition for an A-TGRS code} \label{section 4}
\subsection{Applications of the inverse of the Vandermonde matrix in constructing MDS codes}

 In this section, we find a necessary and sufficient condition such that an A-TGRS code $C$ is MDS. Up to the equivalence of codes [16, Section 2.1], we may assume that $\mathbf{v}=(1,1,\cdots,1)$, i.e. $C=ev_{\alpha,1}(S)$.

It is clear that, $C$ has the generator matrix

\begin{equation}\label{generator matrix}
G=\left(\begin{array}{cccc}
1+\sum\limits_{i=1}^{n-k}\eta_{0,i}\alpha_{1}^{k-1+i} & 1+\sum\limits_{i=1}^{n-k}\eta_{0,i}\alpha_{2}^{k-1+i} & \cdots&1+\sum\limits_{i=1}^{n-k}\eta_{0,i}\alpha_{n}^{k-1+i}\\
\alpha_{1}+\sum\limits_{i=1}^{n-k}\eta_{1,i}\alpha_{1}^{k-1+i}&\alpha_{2}+\sum\limits_{i=1}^{n-k}\eta_{1,i}\alpha_{2}^{k-1+i}&\cdots &\alpha_{n}+\sum\limits_{i=1}^{n-k}\eta_{1,i}\alpha_{n}^{k-1+i}\\
 \vdots&  \vdots& &\vdots\\
   \alpha_{1}^{k-1}+\sum\limits_{i=1}^{n-k}\eta_{k-1,i}\alpha_{1}^{k-1+i} &\alpha_{2}^{k-1}+\sum\limits_{i=1}^{n-k}\eta_{k-1,i}\alpha_{2}^{k-1+i} &\cdots &\alpha_{n}^{k-1}+\sum\limits_{i=1}^{n-k}\eta_{k-1,i}\alpha_{n}^{k-1+i}
\end{array}
\right),
\end{equation}\\
which can be denoted by $G=\left(\begin{array}{cc}
I_{k}& A(\eta)
\end{array}\right)V_{n}(\alpha),$ where $$I_{k}=\left(\begin{array}{cccc}
 1& & & \\
 &1& &  \\
 & &\ddots& \\
 & & & 1
 \end{array}
 \right),\ A(\eta)=\left(\begin{array}{cccc}
 \eta_{0,1}&\eta_{0,2}& \cdots& \eta_{0,n-k}\\
 \eta_{1,1} &  \eta_{1,2}&\cdots & \eta_{1,n-k} \\
 \vdots& \vdots& & \vdots\\
 \eta_{k-1,1}& \eta_{k-1,2}&\cdots & \eta_{k-1,n-1}
 \end{array}
 \right),\ V_{n}(\alpha)=\left(\begin{array}{cccc}
1&1&\cdots&1\\
\alpha_{1}&\alpha_{2}&\cdots&\alpha_{n}\\
\vdots&\vdots&&\vdots\\
\alpha_{1}^{n-1}&\alpha_{2}^{n-1}&\cdots&\alpha_{n}^{n-1}
\end{array}\right).$$   \\

 In fact, $C$ is MDS if and only
if all the minor matrices of order $k$ of $G$ are not zero.

For the convenience of the following description, we give some notations. Let $I$ be any $k$-subset of $\{1,2,\cdots,n\}$. We list the following series of symbols:
\begin{equation}
G(x)=\prod\limits_{ i\in I}(x-\alpha_{i})=c_{0}x^{k}+c_{1}x^{k-1}+\cdots+c_{k-1}x+c_{k},
\end{equation}
\begin{equation}
A_{I}=\left(\begin{array}{cccccc}
0 & 1 & & &\\
 & 0 & 1 &  &\\
  &   & \ddots  &\ddots &\\
  &   &   &     0&    1   &\\
 -c_{k}&-c_{k-1}& -c_{k-2}&\cdots&-c_{1}\\
\end{array}
\right),
\end{equation}
\begin{equation}
d_{j}=c_{k-j}, \ a_{m,t}^{l}=\sum\limits_{
 i+j=l, 1\leq i\leq n-k, 0\leq j\leq t-1}\eta_{m,i}d_{j},
\end{equation}
\begin{equation}
 F_{m,t}(x)=\sum\limits_{l=t}^{n-k+t-1}a_{m,t}^{l}x^{l-t},\  g_{m,t}=-\gamma F_{m,t}(A_{I})\gamma^{T}
\end{equation}
\begin{equation}
M(n,k,\alpha,A(\eta),I)=\left|\begin{array}{cccc}
 1+g_{0,1}&g_{0,2}&\cdots&g_{0,n-k}\\
 g_{1,1}&1+g_{1,2}&\cdots&g_{1,n-k}\\
 \vdots&\vdots&\ddots&\vdots\\
 g_{k-1,1}&g_{k-1,2}&\cdots&1+g_{k-1,n-k}
 \end{array}
 \right|,
\end{equation}

 where $\gamma=(0,\cdots,0,1)$, $j=0,1,\cdots,k$, $m=0,1,\cdots k-1$, $t=1,2,\cdots,k$, $l=1,2,\cdots,n-1$.

\begin{theorem}\label{c1 MDS condition}
Suppose that $3\leq k<n$ and $\alpha_{1},\alpha_{2},\cdots,\alpha_{n}\in \mathbb{F}_{q}$ are distinct elements. Let $\Omega$ be the set of $A(\eta)$ such that
\begin{equation}\label{parameter equation}
M(n,k,\alpha,A(\eta),I)\neq0
\end{equation}
for each $k$-subset $I$ of $\{1,2,\cdots,n\}$. Then the code $C$ constructed by (\ref{generator matrix}) is MDS if and only if $$A(\eta)\in\Omega.$$

\end{theorem}
%\begin{theorem}
%Let $C$ be the TRS code constructed by (\ref{generator matrix}) with different elements $\alpha_{1},\alpha_{2},\cdots,\alpha_{n}\in \mathbb{F}_{q}^{*}$, $A(\eta)=(\eta_{ij})_{k\times(n-k)}\in \mathbb{F}_{q}^{k\times(n-k)}$, as parameters. Then $C$ is  MDS if and only if
%$$M(n,k,\alpha,A(\eta),I)=\left|\begin{array}{cccc}
 %1+g_{01}&g_{02}&\cdots&g_{0,n-k}\\
 %g_{11}&1+g_{12}&\cdots&g_{1,n-k}\\
 %\vdots&\vdots&\ddots&\vdots\\
% g_{k-1,1}&g_{k-1,2}&\cdots&1+g_{k-1,n-k}
% \end{array}
 %\right|\neq0 \ \ \mathrm{for}\ \ all\ \ \ I=\{i_{1},\cdots,i_{k}\}\subseteq\{1,\cdots,n\}.$$

%where $\prod\limits_{i_{j}\in I}(x-\alpha_{i_{j}})=\sum\limits_{j=0}^{k}c_{j}x^{k-j},$ $a_{m,l}=\sum\limits_{i=1}^{l}\eta_{mi}d_{l-i}$,  $F_{m,t}(x)=\sum\limits_{l=t}^{n-k+t-1}a_{m,l}x^{l-t}$, $g_{mt}=-\gamma F_{m,t}(A_{I})\gamma^{T}$, $\gamma=(0,\cdots,0,1)$,
% \ \ (\ By \ \ Theorem\ \ref{State transition formula}\ ),\ let\ \ $F_{m,t}(x)=\sum\limits_{l=t}^{n-k+t-1}a_{m,l}x^{l-t}=\gamma F_{m,t}(A_{I})\gamma^{T}$
%$$A_{I}=\left(\begin{array}{ccccc}
%0 & 1 & & &\\
%0&0&1 & &\\
% \vdots&\vdots&\ddots &\ddots&\\
% 0&0&\cdots &0 &1\\
%-c_{k}&-c_{k-1}&-c_{k-2}&\cdots&-c_{1}\\
% \end{array}
%\right).$$
%\end{theorem}

$\mathbf{Proof.}$ Take any $k$-subset $I$, without loss of generality, we assume $\ I=\{1,2,\cdots,k\}$. Then
$$G(x)=\prod\limits_{1\leq j\leq k}(x-\alpha_{j})=\sum\limits_{j=0}^{k}c_{j}x^{k-j},$$
$G_{h}(x)=\sum\limits_{j=0}^{h}c_{k-j}x^{j}$, $u_{i}=\frac{1}{G^{'}(\alpha_{i})}, i=1,2,\cdots,k$, and $w_{t}=\sum\limits_{j=1}^{k}\alpha_{j}^{t}u_{j}$ for any positive integer $t$.
So, we have the determinant

\begin{align*}
&\left|\begin{array}{cccc}
1+\sum\limits_{i=1}^{n-k}\eta_{0,i}\alpha_{1}^{k-1+i} & 1+\sum\limits_{i=1}^{n-k}\eta_{0,i}\alpha_{2}^{k-1+i} & \cdots&1+\sum\limits_{i=1}^{n-k}\eta_{0,i}\alpha_{k}^{k-1+i}\\
\alpha_{1}+\sum\limits_{i=1}^{n-k}\eta_{1,i}\alpha_{1}^{k-1+i}&\alpha_{2}+\sum\limits_{i=1}^{n-k}\eta_{1,i}\alpha_{2}^{k-1+i}&\cdots &\alpha_{k}+\sum\limits_{i=1}^{n-k}\eta_{1,i}\alpha_{k}^{k-1+i}\\
 \vdots&  \vdots& &\vdots\\
   \alpha_{1}^{k-1}+\sum\limits_{i=1}^{n-k}\eta_{k-1,i}\alpha_{1}^{k-1+i} &\alpha_{2}^{k-1}+\sum\limits_{i=1}^{n-k}\eta_{k-1,i}\alpha_{2}^{k-1+i} &\cdots &\alpha_{k}^{k-1}+\sum\limits_{i=1}^{n-k}\eta_{k-1,i}\alpha_{k}^{k-1+i}
\end{array}\right|\\
  \\
 &=\left|\left(\begin{array}{cccccccc}
 1& & &  &\eta_{0,1}&\eta_{0,2}& \cdots& \eta_{0,n-k}\\
 &1& &  &\eta_{1,1} &  \eta_{1,2}&\cdots & \eta_{1,n-k} \\
 & &\ddots& &\vdots& \vdots& & \vdots\\
 & & & 1  &\eta_{k-1,1}& \eta_{k-1,2}&\cdots & \eta_{k-1,n-1}
 \end{array}
 \right)\left(\begin{array}{cccc}
 1&1&\cdots&1\\
\alpha_{1}&\alpha_{2}&\cdots&\alpha_{k}\\
\vdots&\vdots&&\vdots\\
\alpha_{1}^{n-1}&\alpha_{2}^{n-1}&\cdots&\alpha_{k}^{n-1}
\end{array}\right)\right|\\
& &\\
&=\left|\left(\begin{array}{cc}
I_{k}&A(\eta)\end{array}\right)\left(\begin{array}{c}
V_{k}(\alpha)\\
\overline{V}_{n-k}(\alpha)
\end{array}\right)\right| \ \ where\ \  \overline{V}_{n-k}(\alpha)=\left(\begin{array}{cccc}
 \alpha_{1}^{k}&\alpha_{2}^{k}&\cdots&\alpha_{k}^{k}\\
\alpha_{1}^{k+1}&\alpha_{2}^{k+1}&\cdots&\alpha_{k}^{k+1}\\
\vdots&\vdots&&\vdots\\
\alpha_{1}^{n-1}&\alpha_{2}^{n-1}&\cdots&\alpha_{k}^{n-1}
\end{array}\right) \\
&=\left|I_{k}V_{k}(\alpha)+A(\eta)\overline{V}_{n-k}(\alpha)\right|\\
& &\\
&=\left|I_{k}+A(\eta)\overline{V}_{n-k}(\alpha)(V_{k-1}(\alpha))^{-1}\right||V_{k}(\alpha)|.
\end{align*}

We will discuss the $(m,t)$th entry of the matrix $A(\eta)\overline{V}_{n-k}(\alpha)(V_{k-1}(\alpha))^{-1}$ for two cases.

(1) $\alpha_{1},\alpha_{2},\cdots,\alpha_{n}\in \mathbb{F}_{q}^{*}$.

By Theorem \ref{vandermonde matrix inverse}, we know that

\begin{align*}
&A(\eta)\overline{V}_{n-k}(\alpha)(V_{k-1}(\alpha))^{-1}\ \\
&=-\left(\begin{array}{cccc}
\eta_{0,1}&\eta_{0,2}& \cdots& \eta_{0,n-k}\\
\eta_{1,1} &  \eta_{1,2}&\cdots & \eta_{1,n-k} \\
\vdots& \vdots& & \vdots\\
 \eta_{k-1,1}& \eta_{k-1,2}&\cdots & \eta_{k-1,n-k}
 \end{array}
 \right)\left(\begin{array}{cccc}
\alpha_{1}^{k}&\alpha_{2}^{k}&\cdots&\alpha_{k}^{k}\\
\alpha_{1}^{k+1}&\alpha_{2}^{k+1}&\cdots&\alpha_{k}^{k+1}\\
\vdots&\vdots&&\vdots\\
\alpha_{1}^{n-1}&\alpha_{2}^{n-1}&\cdots&\alpha_{k}^{n-1}
\end{array}\right)\left(\begin{array}{ccc}
\frac{u_{1}}{\alpha_{1}}G_{0}(\alpha_{1})&\cdots&\frac{u_{1}}{\alpha_{1}^{k}}G_{k-1}(\alpha_{1})\\
\frac{u_{2}}{\alpha_{2}}G_{0}(\alpha_{2})&\cdots&\frac{u_{2}}{\alpha_{2}^{k}}G_{k-1}(\alpha_{2}) \\
\vdots&\ddots&\vdots\\
\frac{u_{k}}{\alpha_{k}}G_{0}(\alpha_{k})&\cdots&\frac{u_{k}}{\alpha_{k}^{k}}G_{k-1}(\alpha_{k})
 \end{array}
\right)\\
\end{align*}
\begin{align*}
&=-\left(\begin{array}{cccc}
f_{0}(\alpha_{1})&f_{0}(\alpha_{2})&\cdots&f_{0}(\alpha_{k})\\
f_{1}(\alpha_{1})&f_{1}(\alpha_{2})&\cdots&f_{1}(\alpha_{k})\\
\vdots& \vdots& & \vdots\\
f_{k-1}(\alpha_{1})&f_{k-1}(\alpha_{2})&\cdots&f_{k-1}(\alpha_{k})
\end{array}\right)\left(\begin{array}{ccc}
\frac{u_{1}}{\alpha_{1}}G_{0}(\alpha_{1})&\cdots&\frac{u_{1}}{\alpha_{1}^{k}}G_{k-1}(\alpha_{1})\\
\frac{u_{2}}{\alpha_{2}}G_{0}(\alpha_{2})&\cdots&\frac{u_{2}}{\alpha_{2}^{k}}G_{k-1}(\alpha_{2}) \\
\vdots&\ddots&\vdots\\
\frac{u_{k}}{\alpha_{k}}G_{0}(\alpha_{k})&\cdots&\frac{u_{k}}{\alpha_{k}^{k}}G_{k-1}(\alpha_{k})
 \end{array}
\right),
\end{align*}
where $f_{m}(x)=\sum\limits_{i=1}^{n-k}\eta_{m,i}x^{k-1+i},m=0,1,\cdots,k-1$.

Let $b_{m,t}$ represent the $(m,t)th$ entry of matrix  $A\overline{V}_{n-k}(\alpha)(V_{k-1}(\alpha))^{-1}$. Then
\begin{align*}
b_{m,t}&=-\left(f_{m}(\alpha_{1}),f_{m}(\alpha_{2}),\cdots,f_{m}(\alpha_{k})\right)\left(\begin{array}{c}
\frac{u_{1}}{\alpha_{1}^{t}}G_{t-1}(\alpha_{1})\\
\frac{u_{2}}{\alpha_{2}^{t}}G_{t-1}(\alpha_{2})\\
\vdots\\
\frac{u_{k}}{\alpha_{k}^{t}}G_{t-1}(\alpha_{k})
\end{array}
\right)\\
&\\
&=-f_{m}(\alpha_{1})\frac{u_{1}}{\alpha_{1}^{t}}G_{t-1}(\alpha_{1})+f_{m}(\alpha_{2})\frac{u_{2}}{\alpha_{2}^{t}}G_{t-1}(\alpha_{2})+\cdots+f_{m}(\alpha_{k})\frac{u_{k}}{\alpha_{k}^{t}}G_{t-1}(\alpha_{k})\\
&\\
&=-\left(\eta_{m,1},\eta_{m,2},\cdots,\eta_{m,n-k}\right)\left(\sum\limits_{i=1}^{k}\left(\begin{array}{c}
\alpha_{i}^{k}\\
\alpha_{i}^{k+1}\\
\vdots\\
\alpha_{i}^{n-1}
\end{array}
\right)\frac{u_{i}}{\alpha_{i}^{t}}(1,\alpha_{i},\cdots,\alpha_{i}^{t-1})
\right)\left(\begin{array}{c}
d_{0}\\
d_{1}\\
\vdots\\
d_{t-1}\end{array}
\right)\\
&=-\left(\eta_{m,1},\eta_{m,2},\cdots,\eta_{m,n-k}\right)\left(\begin{array}{cccc}
w_{k-t}&w_{k-t+1}&\cdots&w_{k-1}\\
w_{k-t+1}&w_{k-t+2}&\cdots&w_{k}\\
\vdots&\vdots&&\vdots\\
w_{n-t-1}&w_{n-t}&\cdots&w_{n-2}
\end{array}
\right)\left(\begin{array}{c}
d_{0}\\
d_{1}\\
\vdots\\
d_{t-1}\end{array}
\right)\\
&=-\sum\limits_{l=1}^{n-k+t-1}\sum\limits_{i+j=l; 1\leq i\leq n-k; 0\leq j\leq t-1}\eta_{m,i}d_{j}w_{k-1-t+l}\\
&=-\sum\limits_{l=1}^{n-k+t-1}a_{m,t}^{l}w_{k-1-t+l}\\
&=-\sum\limits_{l=t}^{n-k+t-1}a_{m,t}^{l}w_{k-1-t+l} \ \ ( \mathrm{by \ \ Theorem}\  \ref{wk and diagram},\  w_{0}=w_{1}=\cdots=w_{k-2}=0,w_{k-1}=1\  ).\\
\end{align*}

(2) $0\in \{\alpha_{1},\alpha_{2},\cdots,\alpha_{n}\}$.

If $0 \notin \{\alpha_{i},i\in I\}$, it will be the same condition as  in (1).

Followingly, we consider the condition $0\in\{\alpha_{i},i\in I\}$. Without loss of generality, let $\alpha_{1}=0$.
$$(V_{k-1}(\alpha))=\left(\begin{array}{cccc}
 1&1&\cdots&1\\
 0&\alpha_{2}&\cdots&\alpha_{k}\\
 \vdots&\vdots&\ddots&\vdots\\
 0&\alpha_{2}^{k-1}&\cdots&\alpha_{k}^{k-1}
 \end{array}
 \right)\triangleq \left(\begin{array}{cc}
 1&\beta\\
 0&V_{2}\end{array}\right),$$
  where $V_{2}=\left(\begin{array}{cccc}
 \alpha_{2}&\alpha_{3}\cdots&\alpha_{k}\\
 \alpha_{2}^{2}&\alpha_{3}^{2}\cdots&\alpha_{k}^{2}\\
 \vdots&\vdots&\vdots\\
 \alpha_{2}^{k-1}&\alpha_{3}^{k-1}\cdots&\alpha_{k}^{k-1}
 \end{array}
 \right)$, $0=\left(\begin{array}{c}
 0\\
 0\\
 \vdots\\
 0
 \end{array}
 \right)$ and $\beta=(1,1,\cdots,1).$
 It is not difficult to discover that $$(V_{k-1}(\alpha))^{-1}=\left(\begin{array}{cc}
 1&-\beta V_{2}^{-1}\\
 0&V_{2}^{-1}
 \end{array}\right).$$
 Then by Theorem \ref{vandermonde matrix inverse}, we have
 $$\overline{V}_{n-k}(\alpha)(V_{k-1}(\alpha))^{-1}=\left(\begin{array}{cccc}
 0&\alpha_{2}^{k}&\cdots&\alpha_{k}^{k}\\
 0&\alpha_{2}^{k+1}&\cdots&\alpha_{k}^{k+1}\\
 \vdots&\vdots&&\vdots\\
 0&\alpha_{2}^{n-1}&\cdots&\alpha_{k}^{n-1}
 \end{array}
 \right)\left(\begin{array}{ccccc}
 1&b_{1}&\cdots&b_{k-1}\\
 0&-\frac{u'_{2}}{\alpha_{2}^{2}}g_{0}(\alpha_{2})&\cdots&-\frac{u'_{2}}{\alpha_{2}^{k}}g_{k-2}(\alpha_{2})\\
  \vdots&\vdots&\ddots&\vdots\\
 0&-\frac{u'_{k}}{\alpha_{k}}g_{0}(\alpha_{k})&\cdots&-\frac{u'_{k}}{\alpha_{k}^{k}}g_{k-2}(\alpha_{k})
 \end{array}
 \right)\triangleq\left(\begin{array}{cc}
 0&B
 \end{array}
 \right),$$
 where $g(x)=\prod\limits_{i=2}^{n}(x-\alpha_{i})=\sum\limits_{i=0}^{k-1}d_{i}^{'}x^{i}$, $g_{h}(x)=\sum\limits_{i=0}^{h}d'_{i}x^{i}$, $u'_{i}=\frac{1}{g'(\alpha_{i})},\  h=0,1,\cdots,k-1, \ \ i=2,3,\cdots,k$.

 By Corollary \ref{expression of Vandermonde matrix} and Corollary \ref{zero and wk}, we have
 \begin{align*}
 B&=-\left(\begin{array}{cccc}
 \alpha_{2}^{k}&\alpha_{3}^{k}&\cdots&\alpha_{k}^{k}\\
 \alpha_{2}^{k+1}&\alpha_{3}^{k+1}&\cdots&\alpha_{k}^{k+1}\\
 &\vdots&\vdots&\vdots\\
 \alpha_{2}^{n-1}&\alpha_{3}^{n-1}&\cdots&\alpha_{k}^{n-1}
 \end{array}
 \right)\left(\begin{array}{cccc}
 u'_{2}&&&\\
 &u'_{3}&&\\
 &&\ddots&\\
 &&&u'_{k}
 \end{array}\right)\left(\begin{array}{cccc}
 \frac{1}{\alpha_{2}^{2}}&\frac{1}{\alpha_{2}^{3}}&\cdots&\frac{1}{\alpha_{2}^{k}}\\
 \frac{1}{\alpha_{3}^{2}}&\frac{1}{\alpha_{3}^{3}}&\cdots&\frac{1}{\alpha_{3}^{k}}\\
 \vdots&\vdots&&\vdots\\
 \frac{1}{\alpha_{k}^{2}}&\frac{1}{\alpha_{k}^{3}}&\cdots&\frac{1}{\alpha_{k}^{k}}
 \end{array}
 \right)\left(\begin{array}{cccc}
 d'_{0}&d'_{1}&\cdots&d'_{k-2}\\
 &d'_{0}&\cdots&d'_{k-3}\\
 &&\ddots&\vdots\\
 &&&d'_{0}
 \end{array}\right)\\
  &=-\left(\begin{array}{cccc}
 w'_{k-2}&w'_{k-3}&\cdots&w'_{0}\\
 w'_{k-1}&w'_{k-2}&\cdots&w'_{1}\\
 \vdots&\vdots&&\vdots\\
 w'_{n-3}&w'_{n-4}&\cdots&w'_{n-1-k}
 \end{array}\right)\left(\begin{array}{cccc}
 d'_{0}&d'_{1}&\cdots&d'_{k-2}\\
 &d'_{0}&\cdots&d'_{k-3}\\
 &&\ddots&\vdots\\
 &&&d'_{0}
 \end{array}\right)\\
 \end{align*}
 \begin{align*}
 &=-\left(\begin{array}{cccc}
 w_{k-1}&w_{k-2}&\cdots&w_{1}\\
 w_{k}&w_{k-1}&\cdots&w_{2}\\
 &\vdots&\vdots&\vdots\\
 w_{n-2}&w_{n-3}&\cdots&w_{n-k}
 \end{array}\right)
  \left(\begin{array}{cccc}
 d_{1}&d_{2}&\cdots&d_{k-1}\\
 &d_{1}&\cdots&d_{k-2}\\
 &&\ddots&\vdots\\
 &&&d_{1}
 \end{array}
 \right),
 \end{align*}
 where $w'_{t}=\sum\limits_{i=2}^{n}\alpha_{i}^{t}u'_{i}$ for any integer $t$. Then

 \begin{align*}
  \overline{V}_{n-k}(\alpha)(V_{k-1}(\alpha))^{-1}&=
 \left(\begin{array}{cc}
 0&B
 \end{array}
 \right)\\
 &=-\left(\begin{array}{cccc}
 w_{k-1}&w_{k-2}&\cdots&w_{1}\\
 w_{k}&w_{k-1}&\cdots&w_{2}\\
 \vdots&\vdots&&\vdots\\
 w_{n-2}&w_{n-3}&\cdots&w_{n-k}
 \end{array}\right)
  \left(\begin{array}{ccccc}
 0&d_{1}&d_{2}&\cdots&d_{k-1}\\
 0&&d_{1}&\cdots&d_{k-2}\\
 \vdots&&&\ddots&\vdots\\
 0&&&&d_{1}
 \end{array}\right) \\
 &=-\left(\begin{array}{ccccc}
 w_{k-1}&w_{k-2}&\cdots&w_{1}&w_{0}\\
 w_{k}&w_{k-1}&\cdots&w_{2}&w_{1}\\
 \vdots&\vdots&\vdots&&\vdots\\
 w_{n-2}&w_{n-3}&\cdots&w_{n-k}&w_{n-k-1}
 \end{array}\right)\left(\begin{array}{ccccc}
 0&d_{1}&d_{2}&\cdots&d_{k-1}\\
 0&&d_{1}&\cdots&d_{k-2}\\
 \vdots&&&\ddots&\vdots\\
 &&&&d_{1}\\
 0&0&0&\cdots&0
 \end{array}\right)\\
 &=-\left(\begin{array}{ccccc}
 w_{k-1}&w_{k-2}&\cdots&w_{1}&w_{0}\\
 w_{k}&w_{k-1}&\cdots&w_{2}&w_{1}\\
 \vdots&\vdots&\vdots&&\vdots\\
 w_{n-2}&w_{n-3}&\cdots&w_{n-k}&w_{n-k-1}
 \end{array}\right)\left(\begin{array}{cccc}
 d_{0}&d_{1}&\cdots&d_{k-1}\\
 &d_{0}&\cdots&d_{k-2}\\
 &&\ddots&\vdots\\
 &&&d_{0}
  \end{array}\right).
   \end{align*}
Let $b_{m,t}$ be the $(m,t)$th entry of the matrix $A(\eta)\overline{V}_{n-k}(\alpha)(V_{k-1}(\alpha))^{-1}$. Then
\begin{align*}
b_{m,t}&=-\left(\begin{array}{cccc}
\eta_{m,1}&\eta_{m,2}&\cdots&\eta_{m,n-k}
\end{array}
\right)
\left(\begin{array}{c}
d_{t-1}w_{k-1}+d_{t-2}w_{k-2}+\cdots+d_{0}w_{0}\\
d_{t-1}w_{k}+d_{t-2}w_{k-1}+\cdots+d_{0}w_{1}\\
\vdots\\
d_{t-1}w_{n-2}+d_{t-2}w_{n-3}+\cdots+d_{0}w_{n-k-1}
\end{array}
\right)\\
&=-\sum\limits_{l=1}^{n-k+t-1}\sum\limits_{i+j=l; 1\leq i\leq n-k; 0\leq j\leq t-1}\eta_{m,i}d_{j}w_{k-1-t+l}\\
&=-\sum\limits_{l=1}^{n-k+t-1}a_{m,t}^{l}w_{k-1-t+l}=-\sum\limits_{l=t}^{n-k+t-1}a_{m,t}^{l}w_{k-1-t+l}.
\end{align*}

By Theorem \ref{recursive formulation of wk}, we get
$$\left(\begin{array}{c}
  w_{l-t}\\
  w_{l-t+1}\\
  \vdots\\
  w_{k-1+l-t}
  \end{array}
  \right)=A_{I}^{l-t}\left(\begin{array}{c}
w_{0}\\
 w_{1}\\
 \vdots\\
 w_{k-1}
 \end{array}
 \right),$$
 then
\begin{align*}
\sum\limits_{l=t}^{n-k+t-1}a_{m,t}^{l}\left(\begin{array}{c}
w_{l-t}\\
w_{l-t+1}\\
\vdots\\
w_{k-1+l-t}
\end{array}
\right)=\sum\limits_{l=t}^{n-k+t-1}a_{m,t}^{l}A_{I}^{l-t}\left(\begin{array}{c}
w_{0}\\
w_{1}\\
\vdots\\
w_{k-1}
\end{array}
\right)=\sum\limits_{l=t}^{n-k+t-1}a_{m,t}^{l}A_{I}^{l-t}\left(\begin{array}{c}
0\\
0\\
\vdots\\
1
\end{array}
\right)=\sum\limits_{l=t}^{n-k+t-1}a_{m,t}^{l}A_{I}^{l-t}\gamma.
\end{align*}
By $F_{m,t}(x)=a_{m,t}^{t}+a_{m,t}^{t+1}x+\cdots+a_{m,t}^{n-k+t-1}x^{n-k-1}$. Then we have
 $$b_{m,t}=-\sum\limits_{l=t}^{n-k-1+t}a_{m,t}^{l}w_{k-1+l-t}=-(0,0,\cdots,1)\sum\limits_{l=t}^{n-k+t-1}a_{m,t}^{l}A_{I}^{l-t}\gamma^{T}=-\gamma F_{m,t}(A_{I})\gamma^{T}=g_{m,t}.$$\\
So $C$ is MDS if and only if
$$M(n,k,\alpha,A(\eta),I)=\left|\begin{array}{cccc}
 1+g_{0,1}&g_{0,2}&\cdots&g_{0,n-k}\\
 g_{1,1}&1+g_{1,2}&\cdots&g_{1,n-k}\\
 \vdots&\vdots&\ddots&\vdots\\
 g_{k-1,1}&g_{k-1,2}&\cdots&1+g_{k-1,n-k}
 \end{array}
 \right|\neq0$$
for any $k$-subset $I$.

\begin{corollary} When the matrix $A(\eta)$ takes some special values and  $A(\eta)\in\Omega$, the A-TGRS codes are the corresponding known MDS codes. We list them in the following:
 \begin{itemize}
\item[(1)] When $\eta_{i,j}=0$ except $\eta_{0,1}=\eta\in \mathbb{F}_{q}^{*}$, $C$ is the code discussed in \cite{8} for the case $(t,h)=(1,0)$, i.e,
$$M(n,k,\alpha,A(\eta),I)=\left|\begin{array}{cccc}
 1+g_{0,1}&g_{0,2}&\cdots&g_{0,n-k}\\
 0&1&\cdots&0\\
 \vdots&\vdots&\ddots&\vdots\\
 0&0&\cdots&1
 \end{array}
 \right|=(1+g_{0,1})$$ and $g_{0,1}=-\eta c_{k}$. Therefore $C$ is MDS if and only if $\eta c_{k}\neq1$.

%So Lemma 4 in \cite{8} is a special case of Theorem \ref{c1 MDS condition} in this paper.\\

 \item[(2)]  When $\eta_{i,j}=0$ except $\eta_{k-1,1},\eta_{k-1,2},\cdots,\eta_{k-1,n-k} \in \mathbb{F}_q^{\ast}$, $C$ is the code discussed in \cite{4}.\\

 Using the $k$ minor matrices of $G$, by Corollary \ref{minus Toplitz matirx inverse}, we can get

\begin{eqnarray*}
& &\left|\begin{array}{cccc}
1 & 1 & \cdots&1\\
\alpha_{1}&\alpha_{2}&\cdots &\alpha_{k}\\
\vdots&\vdots& &\vdots\\
\alpha_{1}^{k-2}&\alpha_{2}^{k-2}&\cdots &\alpha_{k}^{k-2}\\
\alpha_{1}^{k-1}+\eta\alpha_{1}^{q-2}&\alpha_{2}^{k-1}+\eta\alpha_{2}^{q-2}&\cdots &\alpha_{k}^{k-1}+\eta\alpha_{k}^{q-2}\\
  \end{array}\right|\\
   &&=(1+\eta w_{-1})\prod\limits_{1\leq i<j\leq k}(\alpha_{j}-\alpha_{i})=(1-\eta \frac{1}{c_{k}})\prod\limits_{1\leq i<j\leq k}(\alpha_{j}-\alpha_{i}), \\
     \end{eqnarray*}
which implies that $C$ is MDS if and only if $\eta \frac{1}{c_{k}}\neq1$.

 %So Theorem 3,(2) for $l=k-1$ i \cite{4}is a special case of Theorem \ref{c1 MDS condition}.\\

\item[(3)] When $\eta_{i,j}=0$ except $\eta_{k-1,1}=\eta\in \mathbb{F}_q^{\ast}$, $C$ is the code discussed in  \cite{6}, i.e.,\\

$$M(n,k,\alpha,A(\eta),I)=\left|\begin{array}{cccc}
 1&0&\cdots&0\\
 0&1&\cdots&0\\
 \vdots&\vdots&\ddots&\vdots\\
 g_{k-1,1}&g_{k-1,2}&\cdots&1+g_{k-1,k}
 \end{array}
 \right|=(1+g_{k-1,k})$$
 and $g_{k-1,k}=-\eta c_{1}$. So $C$ is MDS if and only if $\eta c_{1}\neq1$.

%So Lemma 2.6(1) in \cite{6} is a special case of Theorem \ref{c1 MDS condition}.\\

\item[(4)] When $\eta_{i,j}=0$ except $\eta_{h,t+1}=\eta\in \mathbb{F}_q^{\ast}$, then $C$ is the code discussed in \cite{3}, i.e.,\\
$$M(n,k,\alpha,A(\eta),I)=\left|\begin{array}{ccccc}
 1&&&&\\
 &\ddots&&\\
 &&1&&\\
  g_{h,1}&\cdots&&1+g_{h,h+1}\cdots&g_{h,k}\\
  &&&\ddots&\\
  &&&&\\
  &&&&1
 \end{array}
 \right|=(1+g_{h,h+1})$$
and $g_{h,h+1}=-\eta(\sum\limits_{i=0}^{t}c_{k-h+t-i}w_{k-1+i})$, where $0\leq h\leq k-1$. So $C$ is MDS if and only if $\eta\sum\limits_{i=0}^{t}c_{k-h+t-i}w_{k-1+i}\neq1$.

%So Theorem 3.3 in \cite{3} is a special case of Theorem \ref{c1 MDS condition}.\\

\item[(5)] When $\eta_{i,j}=0$ except $\eta_{k-2,1},\eta_{k-2,2},\eta_{k-1,1},\eta_{k-1,2}\in  \mathbb{F}_q^{\ast}$, $C$ is the code discussed in \cite{1}, i.e.,
\begin{align*}
M(n,k,\alpha,A(\eta),I)&=\left|\begin{array}{ccccc}
 1&&&&\\
 &\ddots&&&\\
 &&1&&\\
  g_{k-2,1}&\cdots&&1+g_{k-2,k-1}&g_{k-2,k}\\
 g_{k-1,1}&\cdots&&g_{k-1,k-1}&1+g_{k-1,k}
 \end{array}
 \right|\\
 &\\
 &=\left|\begin{array}{cc}
 1+g_{k-2,k-1}&g_{k-2,k}\\
 g_{k-1,k-1}&1+g_{k-1,k}
 \end{array}
 \right|\\
 &\\
 &=\left|I_{2}+\left(\begin{array}{cc}
 g_{k-2,k-1}&g_{k-2,k}\\
 g_{k-1,k-1}&g_{k-1,k}
 \end{array}\right)\right|
\end{align*}
 $C$ is MDS if and only if $$M(n,k,\alpha,A(\eta),I)\neq0.$$

%So Theorem 3.3 in \cite{1} is a special case of Theorem \ref{c1 MDS condition}.\\

\item[(6)] When $\eta_{i,j}=0$ except $\eta_{k-l,1},\eta_{k-l+1,2},\cdots,\eta_{k-1,l}\in  \mathbb{F}_q^{\ast}$, $C$ is the code discussed in \cite{7}, i.e.,
\begin{align*}
M(n,k,\alpha,A(\eta),I)&=\left|\begin{array}{ccccccc}
1&&&&&&\\
&\ddots&&&&&\\
&&1&&&&\\
&&& 1+g_{k-l,k-l+1}&g_{k-l,k-l+2}&\cdots&g_{k-l,n-k}\\
 &&&g_{k-l+1,k-l+1}&1+g_{k-l+1,k-l+2}&\cdots&g_{k-l+1,n-k}\\
 &&&\vdots&\vdots&\ddots&\vdots\\
 &&&g_{k-1,k-l+1}&g_{k-1,k-l+2}&\cdots&1+g_{k-1,n-k}
 \end{array}
 \right|\\
 &\\
 &=\left|\begin{array}{cccc}
 1+g_{k-l,k-l+1}&g_{k-l,k-l+2}&\cdots&g_{k-l,k-n-k}\\
 g_{k-l+1,k-l+1}&1+g_{k-l+1,k-l+2}&\cdots&g_{k-l+1,n-k}\\
 \vdots&\vdots&\ddots&\vdots\\
 g_{k-1,k-l+1}&g_{k-1,k-l+2}&\cdots&1+g_{k-1,n-k}
 \end{array}
 \right|.
 \end{align*}

So $C$ is MDS if and only if $$M(n,k,\alpha,A(\eta),I)\neq0.$$

%So Theorem 3.3 in \cite{7} is a special case of Theorem \ref{c1 MDS condition}.\\

\item[(7)]When $\eta_{i,j}=0$ except $\eta_{h_{1},t_{1}},\eta_{h_{2},t_{2}},\cdots,\eta_{h_{l},t_{l}}\in  \mathbb{F}_q^{\ast}$, then $C$ is the code discussed in \cite{10-m}, i.e.,
\begin{align*}
M(n,k,\alpha,A(\eta),I)&=\left|\begin{array}{cccccc}
1&&&&&\\
&\ddots&&&&\\
g_{h_{1},1}&\cdots&1+g_{h_{1},h_{1}+1}&\cdots&\cdots&g_{h_{1},n-k}\\
&&\ddots&&&\\
g_{h_{2},1}&\cdots&\cdots&1+g_{h_{2},h_{2}+1}&\cdots&g_{h_{2},n-k}\\
&&&\ddots&&\\
g_{h_{l},1}&\cdots&\cdots&\cdots&1+g_{h_{l},h_{l}+1}&\cdots g_{h_{l},n-k}\\
&&&&\ddots&\\
&&&&&\\
&&&&&1
 \end{array}
 \right|\\
 &=\left|\begin{array}{cccc}
 1+g_{h_{1},h_{1}+1}&g_{h_{1},h_{2}+1}&\cdots&g_{h_{1},h_{l}+1}\\
 g_{h_{2},h_{1}+1}&1+g_{h_{2},h_{2}+1}&\cdots&g_{h_{2},h_{l}+1}\\
 \vdots&\vdots&\ddots&\vdots\\
 g_{h_{l},h_{1}+1}&g_{h_{l},h_{2}+1}&\cdots&1+g_{h_{l},h_{l}+1}
 \end{array}
 \right|.
  \end{align*}
 So $C$ is MDS if and only if $$M(n,k,\alpha,A(\eta),I)\neq0.$$
  \end{itemize}
\end{corollary}
\begin{theorem}
Let $C$ be the code with the matrix $A(\eta)\in \Omega$. We call the number of non-zero rows (columns) the row (columns) weight of $A(\eta)$, denoted by $\mathrm{RW}(A(\eta))$ and $\mathrm{CW}(A(\eta))$, respectively. If $A(\eta)$ satisfies one of the following conditions, $C$ is an MDS TGRS code which is different from the known corresponding ones.\\
\begin{itemize}
\item[(i)] $\mathrm{RW}(A(\eta))=2$ and $\mathfrak{A}\nsubseteq \{(k-2,1),(k-2,2),(k-1,1),(k-1,2)\}$;\\
\item[(ii)]$\mathrm{RW}(A(\eta))>2$ and $\mathrm{RW}(A(\eta))+\mathrm{CW}(A(\eta))<2|\mathfrak{A}|$,
\end{itemize}
where $\mathfrak{A}=\{(i,j)|\eta_{i,j}\neq0,\ \eta_{i,j}\in A(\eta)\}$ and $|\mathfrak{A}|$ denotes the cardinality of $\mathfrak{A}$.
\end{theorem}
$\mathbf{Proof.}$ Note that
 $$A(\eta)=\left(\begin{array}{cccc}
 \eta_{0,1}&\eta_{0,2}& \cdots& \eta_{0,n-k}\\
\eta_{1,1} &  \eta_{1,2}&\cdots & \eta_{1,n-k} \\
\vdots& \vdots& & \vdots\\
 \eta_{k-1,1}& \eta_{k-1,2}&\cdots & \eta_{k-1,n-1}
 \end{array}
 \right).$$
 In fact in \cite{8,4,6,2,3,9,15,5,1,7,10-m}, $A(\eta)$ takes
$$\left(\begin{array}{cccc}
 \eta&0&\cdots&0 \\
 0& 0& \cdots&0  \\
\vdots&\vdots & &\vdots\\
 0&0 &\cdots & 0
 \end{array}
 \right),
 \left(\begin{array}{cccc}
0 &0&\cdots&0 \\
 \vdots& \vdots& &\vdots  \\
0& 0&\cdots & 0\\
 \eta&0&\cdots & 0
 \end{array}
 \right),\left(\begin{array}{ccc}
 0&\cdots&0 \\
 \vdots & &\vdots  \\
0&\cdots\eta\cdots&0 \\
\vdots&&\vdots\\
0& \cdots&0
 \end{array}
 \right),
 \left(\begin{array}{cccc}
 0&0&\cdots&0 \\
 \vdots&\vdots& &\vdots  \\
\eta_{1}&\eta_{2}&\cdots&0\\
 0&0&\cdots &0
 \end{array}
 \right)  \left(\begin{array}{cccc}
 0&0&\cdots&0 \\
 \vdots&\vdots& &\vdots  \\
\eta_{1,1}&\eta_{1,2}&\cdots&0\\
 \eta_{2,1}&\eta_{2,2}&\cdots &0
 \end{array}
 \right)$$
  $$\left(\begin{array}{cccc}
 0&0&\cdots&0 \\
 \vdots&\vdots & &\vdots  \\
\eta_{h,1}&\eta_{h,2} &\cdots&\eta_{h,n-k} \\
\vdots&\vdots&&\vdots\\
0&0& \cdots&0
 \end{array}
 \right),\left(\begin{array}{ccccc}
 &&& &\\
& & & & \\
\eta_{k-l,1}& & & &\\
 &\eta_{k-l+1,2}& & &\\
 &&\ddots&&\\
 &&&\eta_{k-1,l}&
 \end{array}
 \right),\left(\begin{array}{cccccc}
 &&&&& \\
&\eta_{h_{1},t_{1}} & &&&  \\
& &&&&  \\
 &&\eta_{h_{2},t_{2}}&&&  \\
 &&&&&\\
 &&&\ddots&&\\
 &&&&\eta_{h_{l},t_{l}}&
 \end{array}
 \right),$$
 respectively.

 An A-TGRS code is different from the known ones if and only if $A(\eta)$ is different from the corresponding ones discussed in \cite{8,4,6,2,3,9,15,5,1,7,10-m}. $A(\eta)$ satisfying the condition of one of $(i)$ and $(ii)$ is obviously different from them. According to the solutions of the Eq.(\ref{parameter equation}), $A(\eta)$ satisfying the condition $(i)$ or $(ii)$ do exist. The result follows.

\begin{example}
Let $q=11$ and $\alpha=(1,2,3,5,6,8,9,10)$. $C$ is an $[8,k,9-k]$ MDS code if and only if for each $k$-subset $I\subseteq\{1,2,\cdots,n\}$, $M(n,k,\alpha,A(\eta),I)\neq0$ holds. By MATLAB, there are so many $A(\eta)$'s such that the conditions $(i)$ or $(ii)$ hold that we can't list them all. Here we only list one example for each dimension $k$ when $k=3,4,5,6,7$.

\begin{center}
\begin{tabular}{|c|c|c|c|c|c|}
\hline
dimension $k$ & $3$&$4$&$5$&$6$&$7$ \\
\hline
parameter matrix $A(\eta)$ & $\left(\begin{array}{ccccc}
0&0&0&1&10\\
0&0&0&0&7\\
0&0&0&0&0
\end{array}\right)$ & $\left(\begin{array}{cccc}
0&0&0&4\\
0&0&0&7\\
0&0&0&3\\
0&0&2&6
\end{array}\right)$& $\left(\begin{array}{ccc}
0&0&0\\
0&10&0\\
0&9&8\\
0&0&0\\
0&0&0
\end{array}\right)$&$\left(\begin{array}{cc}
7&1\\
4&3\\
0&0\\
0&0\\
0&0\\
0&0
\end{array}\right)$&$\left(\begin{array}{c}
0\\
0\\
0\\
0\\
4\\
6\\
10
\end{array}
\right)$
\\
\hline
\end{tabular}
\end{center}
\end{example}
\subsection{Application of the inverse of the Toplitz matrix}
 As an important application, we give a simplified version of the parity-check matrix of the TGRS code constructed in \cite{4} in this subsection.

 For ease of understanding, we use the same symbols as that in \cite{4}.
Let $\alpha=(\alpha_{1},\alpha_{2},\cdots,\alpha_{n})$ and $v=(v_{1},v_{2},\cdots,v_{n})\in \left(\mathbb{F}_{q}^{*}\right)^{n}$ as the above. Suppose that $C_{1}$ is the $[n,k]$ linear code over $\mathbb{F}_{q}$ with generator matrix
$$G_{k-1}(\alpha,v)=\left(\begin{array}{cccc}
v_{1} & v_{2} & \cdots&v_{n}\\
 v_{1}\alpha_{1}&v_{2}\alpha_{2}&\cdots &v_{n}\alpha_{n}\\
 \vdots&  \vdots& &\vdots\\
 v_{1}(\alpha_{1}^{k-1}+\eta\alpha_{1}^{q-2})& v_{2}(\alpha_{2}^{k-1}+\eta\alpha_{2}^{q-2}) &\cdots &v_{n}(\alpha_{n}^{k-1}+\eta\alpha_{n}^{q-2})
\end{array}
\right).$$
Then $C_{1}$ is a twisted generalized RS (TGRS) code and the parity-check matrix of $C_{1}$ has been discussed in \cite{4}. Denote $G(x)=\prod\limits_{i=1}^{n}(x-\alpha_{i})$, $u_{i}=\frac{1}{G'(\alpha_{i})},i=1,2,\cdots,n$.
\begin{lemma}(Theorem 1 in \cite{4}) \label{c3 parity check matrix}
Let $b_{0}=1$ and $b_{1},b_{2},\cdots,b_{k-l-1}\in \mathbb{F}_{q}$ be given by the following recursion
$$b_{j}=-\frac{\sum\limits_{r=0}^{j-1}\sum\limits_{i=1}^{n}u_{i}\alpha_{i}^{n+j-1-r}}{\sum\limits_{i=1}^{n}u_{i}\alpha_{i}^{n-1}},j=1,2,\cdots,k-l-1.$$
Then the TGRS code $C_{1}$ has a parity-check matrix
$$\left(\begin{array}{c}
\underline{\frac{u_{1}}{v_{1}}f(\alpha_{1})\ \ \frac{u_{2}}{v_{2}}f(\alpha_{2})\ \ \cdots\ \ \frac{u_{n}}{v_{n}}f(\alpha_{n})}\\
G_{n-k-1}(\alpha,u\bigodot v^{-1}\bigodot \alpha)
\end{array}
\right),$$
$\ $
where $u\bigodot v^{-1}\bigodot \alpha=(\frac{u_{1}}{v_{1}}\alpha_{1},\frac{u_{2}}{v_{2}}\alpha_{2},\cdots,\frac{u_{n}}{v_{n}}\alpha_{n})$ and
$f(x)=x^{n-l-1}+b_{1}x^{n-l-2}+\cdots+b_{k-l-1}x^{n-k}-\frac{\sum\limits_{i=1}^{n}u_{i}\alpha_{i}^{n-1}}{\eta \sum\limits_{i=1}^{n}u_{i}\alpha_{i}^{-1}}$.
\end{lemma}
Now we simplify the result in \cite{4} as follows.

\begin{theorem}
Let $G(x)=\prod\limits_{i=1}^{n}(x-\alpha_{i})=\sum\limits_{j=0}^{n}c_{j}x^{n-j}$. Then TGRS code $C_{1}$ has a parity-check matrix
$$\left(\begin{array}{c}
\underline{\frac{u_{1}}{v_{1}}f(\alpha_{1})\ \ \frac{u_{2}}{v_{2}}f(\alpha_{2})\ \ \cdots\ \ \frac{u_{n}}{v_{n}}f(\alpha_{n})}\\
G_{n-k-1}(D,u\bigodot v^{-1}\bigodot \alpha)
\end{array}
\right),$$
 where $f(x)=x^{n-l-1}+c_{1}x^{n-l-2}+\cdots+c_{k-l-1}x^{n-k}-\frac{c_{n}}{\eta}$.
\end{theorem}
$\mathbf{Proof.}$
Suppose $$f(x)=b_{0}x^{n-1-l}+b_{1}x^{n-l-2}+\cdots+b_{k-l-1}x^{n-k}+b\in C_{1}^{\bot}.$$
According to the proof of Theorem 1 in \cite{4}, $b_{0},b_{1},\cdots,b_{k-l-1}$ satisfying the following equation\\
$$\left\{\begin{array}{c}
b_{0}\sum\limits_{i=1}^{n}u_{i}\alpha_{i}^{n}+b_{1}\sum\limits_{i=1}^{n}u_{i}\alpha_{i}^{n-1}=0,\\
b_{0}\sum\limits_{i=1}^{n}u_{i}\alpha_{i}^{n+1}+b_{1}\sum\limits_{i=1}^{n}u_{i}\alpha_{i}^{n}+b_{2}\sum\limits_{i=1}^{n}u_{i}\alpha_{i}^{n-1}=0,\\
\vdots \\
b_{0}\sum\limits_{i=1}^{n}u_{i}\alpha_{i}^{n+k-l-2}+b_{1}\sum\limits_{i=1}^{n}u_{i}\alpha_{i}^{n+k-l-3}+\cdots+b_{k-l-1}\sum\limits_{i=1}^{n}u_{i}\alpha_{i}^{n-1}=0,\\
b_{0}\sum\limits_{i=1}^{n}u_{i}\alpha_{i}^{n-1}+b\eta \sum\limits_{i=1}^{n}u_{i}\alpha_{i}^{-1}=0,
\end{array}\right.$$

which is equivalent to

$\left(\begin{array}{cccc}
w_{n-1}&0&\cdots&0\\
w_{n}&w_{n-1}&\cdots&0\\
\vdots&\ddots& &\\
w_{n+k-l-2}&\cdots&\cdots&w_{n-1}
\end{array}
\right)\left(\begin{array}{c}
b_{0}\\
b_{1}\\
\vdots\\
b_{k-l-1}
\end{array}\right)=\left(\begin{array}{c}
1\\
0\\
\vdots\\
0
\end{array}
\right).$\\

 By Theorem \ref{inverse of  Toplitz Matrix} and Corollary \ref{generalized Toplitz matirx inverse}, we obtain
\begin{align*}
\left(\begin{array}{c}
b_{0}\\
b_{1}\\
\vdots\\
b_{k-l-1}
\end{array}\right)&=\left(\begin{array}{cccc}
w_{n-1}&0&\cdots&0\\
w_{n}&w_{n-1}&\cdots&0\\
\vdots&\ddots& &\\
w_{n+k-l-2}&\cdots&\cdots&w_{n-1}
\end{array}
\right)^{-1}\left(\begin{array}{c}
1\\
0\\
\vdots\\
0
\end{array}
\right)\\
&=\left(\begin{array}{cccc}
1&0&\cdots&0\\
c_{1}&1&\cdots&0\\
\vdots&\ddots& &\\
c_{k-l-1}&\cdots&\cdots&1
\end{array}
\right)\left(\begin{array}{c}
1\\
0\\
\vdots\\
0
\end{array}
\right)=\left(\begin{array}{c}
1\\
c_{1}\\
\vdots\\
c_{n-l-1}
\end{array}
\right).
\end{align*}
From Theorem \ref{wk and diagram}, by direct calculation, we can see $\sum\limits_{i=1}^{n}u_{i}\alpha_{i}^{-1}=(-1)^{n-1}\frac{1}{\alpha_{1}}\frac{1}{\alpha_{2}}\cdots \frac{1}{\alpha_{n}}=-\frac{1}{c_{n}}$ and $\sum\limits_{i=1}^{n}u_{i}\alpha_{i}^{n-1}=1$. The result follows.
\section{Conclusion}
Our work mainly includes three parts. Firstly, we give an explicit expression of the inverse of the Vandermonde matrix employing the transitional matrix between two bases of the linear space $\mathbb{F}_{q}[x]_{n}$ over $\mathbb{F}_{q}$. As a consequence of the inverse of the Vandermonde matrix, we get the analyze expression of the inverse of the Toplitz matrix. Furthermore, we also determine the value of each entry of the inverse of the Toplitz matrix by a LFSR sequence. Secondly, as another important application of the inverse of the Vandermonde matrix, we characterized a sufficient and necessary condition under which an A-TGRS code is MDS. According to this condition,  we not only can get all the known MDS TGRS codes in \cite{8,4,6,2,3,9,15,5,1,7,10-m}, but also obtain more MDS TGRS codes with new parameter matrices.
Finally, as an application of the inverse of the Toplitz matrix, we simplify the result in \cite{4}. In fact, it should be pointed out that the applications of the inverses of the Vandermonde matrix and the Toplitz matrix are not limited to the items listed in this paper.

\end{document}